\documentclass[graybox]{svmult}

\usepackage{type1cm}        
\usepackage{makeidx}         
\usepackage{graphicx}        
\usepackage{multicol}        
\usepackage[bottom]{footmisc}

\usepackage{newtxtext}       %
\usepackage[varvw]{newtxmath}       

\def\ba{\begin{eqnarray}}
\def\ea{\end{eqnarray}}

\def\q{\quad}

\makeindex             

\begin{document}

\title*{Spin foams, Refinement limit and Renormalization}
\author{Seth K. Asante, Bianca Dittrich and Sebastian Steinhaus}

\institute{Seth K. Asante \at Theoretisch-Physikalisches Institut, Friedrich-Schiller-Universit\"at Jena, Max-Wien-Platz 1, 07743 Jena, Germany, \email{seth.asante@uni-jena.de}
\and
Bianca Dittrich \at Perimeter Institute f. Theoretical Physics, 31 Caroline St. N, N2L 2Y5 Waterloo, Ontario, Canada, \email{bdittrich@perimeterinsitute.ca}
\and Sebastian Steinhaus \at Theoretisch-Physikalisches Institut, Friedrich-Schiller-Universit\"at Jena, Max-Wien-Platz 1, 07743 Jena, Germany, \email{sebastian.steinhaus@uni-jena.de}}
%
%
\maketitle



\abstract{Spin foams provide path integrals for quantum gravity, which employ discretizations as regulator. To obtain regulator independent predictions, we must remove these fiducial structures in a suitable refinement limit. In this chapter we present the current state of research: We begin with a discussion on the role of diffeomorphism symmetries in discrete systems, the notion of scale in background independent theories and how we can consistently improve theories via renormalization to reduce regulator dependence. 
We  present the consistent boundary formulation, which provides a renormalization framework for background independent theories, and discuss tensor network methods and restricted spin foams, which provide concrete renormalization algorithms aiming at the construction of consistent boundary amplitudes for spin foams.
We furthermore discuss effective spin foams, which have allowed for the construction of a perturbative refinement limit and an associated effective continuum action.}

\section*{Keywords} 
Quantum Gravity, Spin Foams, Regge Calculus, Renormalization, Numerical Methods, Tensor Networks

\section{Introduction}
\label{sec:1}


Spin foams construct non-perturbative path integrals for quantum gravity, which implement a rigorous notion of quantum geometry. Such path integrals require however a regularization, which for spin foams is implemented via a discretization. 

This regulator has to be removed in order to obtain sensible amplitudes, which do not depend on arbitrary discretization choices. That is, we have to determine the refinement limit. This can be implemented via renormalization algorithms, which organize this enormous computational challenge into iterative steps and allow for truncation schemes.

This chapter is organized as follows:
\begin{description}
	\item[\textbf{Sec.~\ref{Sec:basics}}] provides some basic background on Regge calculus and spin foams needed for the following discussions.
	
	\item[\textbf{Sec.~\ref{Sec:need}:}] discusses why constructing the refinement limit is indispensable in order to restore diffeomorphism symmetry and to obtain discretization independent amplitudes.
	
	\item[\textbf{Sec.~\ref{Sec:coupling}:}] explains how to define a notion of scale and renormalization flow of couplings for background independent theories such as spin foams.
	
	\item[\textbf{Sec.~\ref{Sec:CBF}:}] sketches the consistent boundary formalism. This provides a  framework to construct and express the refinement limit via a family of consistent amplitudes for spacetime regions with finer and finer boundary data. The consistency conditions motivate an iterative renormalization scheme, applicable to background independent theories and preserving the locality of (spin foam) amplitudes.

		\item[\textbf{Sec.~\ref{Sec:TNR}:}] discusses tensor network algorithms and their adaption and application to (analogue) spin foams. Tensor network algorithms can be understood as realizing the construction of consistent boundary amplitudes.
		
			\item[\textbf{Sec.~\ref{Sec:Restr}:}] discusses the renormalization flow of restricted spin foams, which arises from a combination of symmetry reduction and semi-classical approximation from the full spin foam models.
			
				\item[\textbf{Sec.~\ref{Sec:ESFs}:}] sketches effective spin foam models and first results on their (perturbative) refinement limit.
	
	\item[\textbf{Sec.~\ref{Sec:Schluss}:}] provides a short summary and comments on some open issues.
\end{description} 

\section{Basics of Regge calculus and spin foams}\label{Sec:basics}

Here we will introduce basic facts and notions of Regge calculus and spin foams.

\textit{Regge calculus} \cite{Regge:1961px} is a discretization of general relativity, defined on triangulations. Its fundamental degrees of freedom are the edge lengths of the triangulation. It is a coordinate-free theory, as we only refer to the distances between vertices, not how these vertices are embedded. The basic building blocks of a $d$-dimensional triangulation are $d$-simplices, which are defined as the convex hull of $(d+1)$ vectors in $\mathbb{R}^d$ (or $\mathbb{R}^{(1,d-1)}$ for Lorentzian signature). So, $d$-simplices can be embedded into $d$-dimensional flat spacetime and are thus intrinsically flat. 
A simplex is uniquely described (up to translations, rotations etc.) by its edge lengths, e.g. six edge lengths for a 3-simplex (tetrahedron) or ten edge lengths for a 4-simplex. From the edge lengths of a simplex we compute its dihedral angles, i.e. the angles enclosed between two $(d-1)$-simplices. The angle is associated to the $(d-2)$-simplex shared by the $(d-1)$-simplices.
Curvature in Regge calculus is distributional and encoded in deficit angles associated to $(d-2)$-simplices, e.g. triangles in 4D. 
A deficit angle measures the difference between $2\pi$ and the sum of dihedral angles of $d$-simplices that contain the $(d-2)$-simplex. 
The dynamics of the theory is encoded in the Regge action, which is a discretization of the Einstein-Hilbert action. 

\textit{Area Regge calculus} \cite{Barrett:1997tx} on the other hand, is a 4D theory that uses areas of triangles instead of edge lengths as its fundamental variables. The $4$-simplices are again intrinsically flat and determined by the values of its ten triangle areas, i.e. the deficit angles are functions of these areas. 
While the number of triangles is the same as the number of edges in a 4-simplex, for larger triangulations we have (generically) much more triangles than edges. 
 To formulate a theory in terms of area variables equivalent to (length) Regge calculus, additional constraints must be imposed \cite{Dittrich:2008va}.

\textit{Spin foams} \cite{Baez:1999sr,Perez:2012wv} are path integrals for quantum gravity, defined on a 2-complex. This 2-complex is frequently chosen to be dual to a triangulation. It consists of vertices, edges and faces. Vertices are dual to $d$-simplices, edges are dual to $(d-1)$-simplices and faces are dual to $(d-2)$-simplices. The 2-complex is colored with group\footnote{The groups in questions are SO$(4)$ for Euclidean signature models and SO$(3,1)$ for Lorentzian signature models.} theoretic data, which are quantum numbers encoding geometry. In 4D, irreducible representations of the group, associated to the faces, encode their areas, whereas tensors invariant under the action of the group, called intertwiners, are assigned to edges. They encode the shape of 3D polyhedra \cite{Baez:1999tk}. A spin foam model then locally assigns amplitudes to the 2-complex: vertex amplitudes to vertices, edge amplitudes to edges and face amplitudes to faces. These amplitudes are derived from general relativity expressed as a constrained topological quantum field theory \cite{Plebanski:1977zz}. A frequently derived result elucidates \cite{Barrett:1998gs, Baez:2002rx, Conrady:2008mk, Barrett:2009gg} the connection of spin foams to (Area) Regge calculus: in the limit where all representations are large, the spin foam vertex amplitude is dominated by boundary data corresponding to geometric $4$-simplices and it oscillates with the Regge action associated to the $4$-simplex.

\section{The need for renormalization: restoring diffeomorphism symmetry}\label{Sec:need}

The spin foam approach, much like Regge gravity \cite{Regge:1961px}, is based on a discretization of the underlying spacetime manifold. Such discretizations do often interfere with the symmetries of the system, in this case diffeomorphism symmetry. 

The question whether Regge gravity features a notion of diffeomorphism symmetry has been debated in the literature \cite{Rocek:1982fr,Loll:1998aj,Hamber:1996pj,Morse:1991te}. On the one hand linearized diffeomorphisms have been identified in linearized Regge gravity around a flat background \cite{Rocek:1982fr}.  In this context the diffeomorphism modes are essential to obtain the correct number of two propagating graviton modes in the continuum limit.  Arguments have been put forward \cite{Rocek:1982fr, Hamber:1996pj}, that the existence of such gauge symmetry modes should extend to the full theory. On the other hand, the data associated to Regge triangulations do not include any obvious notion of embedding coordinates\footnote{We will see below, that this is similar to  time reparametrization invariant systems, where the auxiliary time parameter also does not appear in the description of the discrete dynamics.}, on which diffeomorphism could act. Therefore it has been often suggested that each distinct Regge configuration does define a physically different geometry. Furthermore, Regge calculus solutions with curvature are typically uniquely specified by their boundary data, which speaks against the existence of gauge symmetries. 

This debate arose due to the common, but in this context misleading, characterization of gauge symmetries as (local) transformations acting on the dynamical variables, which leave the action invariant \cite{Hamber:1996pj}. Requiring invariance of the action constitutes only one condition for the action. If the action is a function of $N$ variables, then one can find, locally around each point with non-vanishing gradient, $(N-1)$ such transformations.

The crux of the matter is, however, that most of these transformations act trivially on configurations that define solutions of the system \cite{Dittrich:2008pw,Bahr:2009ku}. In fact, whether gauge symmetries ``exist" (have a non-trivial action) or ``do not exist" (have a trivial action), does depend on the given solution. In the case of Regge gravity (without a cosmological constant) flat solutions do feature gauge symmetries, whereas solutions with curvature typically do not. 

It is indeed advantageous to think of 4D\footnote{
In 3D Regge gravity (without cosmological constant) all solutions are flat and diffeomorphism symmetry is not broken.
}
Regge gravity as a system with diffeomorphism symmetry, but where this symmetry is typically broken \cite{Dittrich:2008pw,Bahr:2009ku,Asante:2018wqy}. This can be turned into a quantitative statement: If there is a non-trivial action of a gauge symmetry on a given solution, then the Hessian of the action, evaluated on this solution, has to have null modes, that is vanishing eigenvalues. Broken symmetries are characterized by non-vanishing eigenvalues, whose absolute value is however much smaller than the absolute value of the remaining eigenvalues. This opens the questions, of how to find discretizations, where these eigenvalues are actually zero, and thus (a discrete version of) diffeomorphism symmetry is restored.

4D Regge gravity solutions with curvature \cite{Bahr:2009ku}, 3D Regge gravity solutions with cosmological constant (and therefore curvature) \cite{Dittrich:2008pw}, as well as Area Regge solutions with non-metricity \cite{Asante:2018wqy} all feature broken diffeomorphism symmetries in this sense. 

The same concepts apply to simple mechanical systems: by adding time $t$ to the configuration variables $q^B$ one can introduce reparametrization invariance as a gauge symmetry: Starting from a  Lagrangian $L(q^B,dq^B/dt)$, we define a new Lagrangian 
 \ba
 L_{ri}(q^B,dq^B/ds,t,dt/ds)=L(q, \frac{dq^B/ds}{dt/ds}) dt/ds  \q ,
\ea 
 with an auxiliary parameter $s$. The action $\int_{s_0}^{s_1} L_{ri} ds$ along a path $(q(s),t(s))$ remains invariant, if we apply a reparametrization $(q(f(s)),t(f(s)))$, where $f$ is a monotonically growing function with $f(s_0)=s_0$ and $f(s_1)=s_1$. This holds also for paths that extremize the action, we have therefore a non-trivial action of reparametrizations on solutions, and therefore a gauge symmetry.
 
 Discretizing such systems one typically breaks reparametrization symmetry  \cite{Bahr:2009mc,Bahr:2011uj}, and loses energy as a constant of motion. The latter is a well-known byproduct of discretizing mechanical systems, and can only be avoided by a  perfect discretization \cite{marsden_west_2001,Bahr:2009mc}. 

Such perfect\footnote{
This notion is inspired by perfect actions developed for Lattice QCD \cite{Hasenfratz:1993sp}.
} discretizations are defined to exactly mirror the continuum dynamics \cite{Bahr:2009qc,Bahr:2011uj}. To be more precise, given a map from the set of  variables in the discretization  to observables in the continuum system, a discretization is said to be perfect, if the discrete solutions do agree with the continuum solutions, as probed by the observables. 

For  mechanical systems (with or without reparametrization invariance), the classical perfect action, that is the action describing a perfect discretization, agrees with the Hamilton-Jacobi function, aka Hamilton's principal function \cite{marsden_west_2001,Bahr:2009qc,Bahr:2009mc}.

That is, consider a dynamical system with time parameter\footnote{
We allow here systems with and without reparametrization invariance. For the latter $s$ can be identified with $t$, for the former $t$ is part of the variables, e.g. $x^0=t$.
}
$s$ and  variables $x^A$. The dynamics is described by a Lagrangian
$
L_{\rm C}(x^A, \dot{x}^A)$,
where $\dot{x}^A$ is $d x^A/ds$. The Hamilton-Jacobi function is defined as action evaluated on solutions, that is 
\ba\label{SHJ}
S_{\rm HJ}(x^A_1,s_1;x^A_0,s_0) &=&\int L_{\rm C}(x^A_{|{\rm sol}}(s), \dot{x}_{|{\rm sol}}^A(s))\,  {\rm d}s \q ,
\ea
where $x^A_{|{\rm sol}}$ extremizes the action, with $x^A_{|{\rm sol}}(s_0)=x^A_0$ and $x^A_{|{\rm sol}}(s_1)=x^A_1$. From this definition follows the following important additivity property
\ba\label{AddHJ}
S_{\rm HJ}(x^A_1,s_1;x^A_0,s_0) &=& \text{Extr}_{x^A_z} \left(S_{\rm HJ}(x^A_1,s_1;x^A_z,s_z)+ S_{\rm HJ}(x^A_z,s_z;x^A_0,s_0) \right)\q .
\ea
Here $\text{Extr}_{x^A_z}F(x^A_z)$ means that we take the extremum of $F$ over the range of $\{x^A_z\}_A$. Note that this extremum is attained at $x^A_{|{\rm sol}}(s_z)$, that is the solution evaluated at $s_z$.

The action for a perfect discretization  is given by
\ba\label{PerfA1}
S_{\rm Per}(\{x^A_n\})&=& \sum_{n=0}^{N-1} S_{\rm HJ}(x^A_{n+1},s_{n+1};x^A_n,s_n)  \q . 
\ea
Here  $\{s_n\}_{n=0}^N$ is a choice of discrete time values  and $\{x^A_n\}_A$ the corresponding set of variables at time parameter $s_n$. This perfect action reproduces the dynamics of the continuum action in the following sense:  Given boundary values $(x_0^A,s_0)$ and $(x_N^A,s_N)$, the solutions, given by the extrema for the discrete action, are given by $x^A_n=x^A_{|{\rm sol}}(s_n)$, where $x^A_{|{\rm sol}}(s)$ is the continuum solution. See \cite{marsden_west_2001} 
for an explicit proof.

With the perfect mirroring of the continuum dynamics by the perfect discretization, one also expects that  symmetries of the continuum action, that can be captured by the discrete variables, are present for the perfect discretization. 

In particular, consider a continuum action with time reparameterization invariance. That is, if for fixed $x^A_0=x^A(s_0)$ and $x^A_N=x^A(s_N)$ we have a solution $x^A_{|{\rm sol}}(s)$, then $x^A_{|{\rm sol}}(f(s))$ , for a monotonically increasing function $f:{\mathbb R}\rightarrow {\mathbb R}$, is also a solution. 
The perfect action (\ref{PerfA1}) will show a discrete mirroring of this symmetry: for fixed boundary values $x^A_0,x^A_N$ an entire family of solutions is given by $x^A_n=x^A_{|{\rm sol}}(f(s_n))$.

Another property, that the perfect action (\ref{PerfA1}) enjoys, is a  notion of discretization independence: The value  of the perfect action (\ref{PerfA1}) evaluated on a solution with given boundary values $(x^A_{\rm in},s_{\rm in}) =(x^A_0,s_0),(x^A_{\rm out},s_{\rm out})=(x^A_N,s_N)$ does not depend on the number of subdivision points, that is $N$. It is always given by the Hamilton-Jacobi function of the continuum system. 

Thus, the perfect action (\ref{PerfA1})  is also a fixed point of the following refinement\footnote{We  use here refinement flow instead of the more widely used term coarse graining flow, as we  first refine and then solve for the degrees of freedom added by the refinement. We will discuss in Section \ref{Sec:coupling} the notion of scale in more detail.} flow: starting from a given discrete action $S_{\rm D}(x^A_1,s_1;x^A_0,s_0)$ for a time interval, we subdivide this interval into two, $(s_1,s_z)$ and $(s_z,s_0)$,  and compute a new action
\ba\label{CG1}
S'_{\rm D}(x^A_1,s_1;x^A_0,s_0)&=& \text{Extr}_{x^A_{\rm z}}  \left(S_{\rm D}(x^A_1,s_1;x^A_z,s_z)+ S_{\rm D}(x^A_z,s_z;x^A_0,s_0) \right)\, .
\ea
 The additivity property (\ref{AddHJ}) of the Hamilton-Jacobi function ensures
that the perfect action (\ref{PerfA1}) is invariant under this flow, $S'_{\rm Per}=S_{\rm Per}$.  

This suggest two ways to calculate a perfect discretization: $(a)$ one can solve the continuum dynamics, and compute the Hamilton-Jacobi function, which then defines the perfect action. This procedure can be described as blocking from the continuum \cite{Bietenholz:1995cy, Bahr:2010cq,Asante:2021blx}, but might be only practical for free theories. Another possibility is to start with a discrete action and subject it to a refinement
flow as described in (\ref{CG1}). Alternatively, one can solve the corresponding fixed point conditions. The latter method allows to find analytical expressions for perfect actions also in the case of more complicated systems \cite{Bahr:2011uj,Asante:2021blx}.

The notion of perfect discretization and refinement flow applies also to the quantum case \cite{Bahr:2011uj,Asante:2021blx}.  The role of the Hamilton-Jacobi function is taken over by the quantum mechanical propagator
\ba\label{Prop1}
K(x^A_1,s_1;x^A_0,s_0)&=& \langle x^A_1,s_1\,| x^A_0,s_0\rangle  \q ,
\ea
which can be defined from the path integral. The additivity property (\ref{AddHJ}) translates into the convolution property for the (continuum) propagator $K_{\rm C}$
\ba\label{Conv1}
K_{\rm C}(x^A_1,s_1;x^A_0,s_0)&=& \int K_{\rm C}(x^A_1,s_1;x^A_z,s_z)K_{\rm C}(x^A_z,s_z;x^A_0,s_0) \,\, \prod_A {\rm d}x_z^A \q .
\ea

We can define a quantum version of the refinement flow (\ref{CG1}), given by
\ba\label{CG2}
K_{\rm D}'(x^A_1,s_1;x^A_0,s_0)&=& \int K_{\rm D}(x^A_1,s_1;x^A_z,s_z)K_{\rm D}(x^A_z,s_z;x^A_0,s_0) \,\, \prod_A {\rm d}x_z^A \q ,
\ea
where $K_{\rm D}$ is a given discretization of the quantum amplitude. The convolution property (\ref{Conv1}) of the continuum amplitude $K_{\rm C}$ ensures, that it is a fixed point of the flow defined by (\ref{CG2}). 
Such fixed points can be obtained by applying (\ref{CG2}) iteratively.
Alternatively one can directly solve the fixed point conditions. The latter method does allow to construct propagators for e.g. the anharmonic oscillator \cite{Bahr:2011uj} and has been used to determine one-loop measures for the path integral \cite{Asante:2021blx}.

We so far assumed general mechanical and quantum mechanical examples in (\ref{SHJ}) and (\ref{Prop1}), with or without a notion of time-reparametrization invariance. If the continuum system do show time-reparametrization invariance, then the  Hamilton-Jacobi function (\ref{SHJ}) and the continuum propagator (\ref{Prop1}) will not depend on the auxiliary\footnote{Often, the variables $x^A$ will include a `physical' time parameter, thus the $s$--parameter is auxiliary.} time parameter values $s_0$ and $s_1$.  
Discretizing such systems one will find that, in most cases, even non-perfect discrete actions and non-perfect discrete propagators  will {\it not} depend on such auxiliary parameters. 

Consider, for example, a reparametrization invariant system, describing a particle in a potential, with variables $x^0=t,x^1=q$. The continuum action is given by
\ba\label{SCRep}
S_{\rm C}&=&\int \!\! ds  \left(  \tfrac{1}{2} \dot q^2 (\dot t)^{-1} - V(q) \,\dot t \right)  \q ,
\ea
which can be discretized to 
\ba\label{SDRep}
S_{\rm D}&=& \sum_{n=0}^{N-1}\tfrac{1}{2}
 \left( 
 (q_{n+1}-q_n)^2 (t_{n+1}-t_n)^{-1} -\tfrac{1}{2}(V(q_{n+1})+V(q_n))\, (t_{n+1}-t_n) 
 \right) \,.\q\q
\ea
The reparametrization invariance is typically
\footnote{
The symmetry is not broken for vanishing potential, as with the discretization (\ref{SDRep}) the particle trajectory is piecewise linear  and the free particle trajectory is linear in $(t,q)$ space.
}
broken by the discretization, i.e. fixing boundary values $(q_N,t_N,q_0,t_0)$, the solutions for the $q_n$ and $t_n$ are unique . 

The $s$-parameter can be understood as auxiliary time coordinate. Its non-appearance in the discretization of reparametrization invariant systems is completely analogous to the non-appearance of spacetime coordinates in Regge calculus. We showed that there is nevertheless a notion of (broken) reparametrization symmetry for these systems. The same holds for (broken) diffeomorphisms  in the case of Regge calculus.

Similar concepts and strategies apply to the discretization of (quantum) field theories.  There is however an important difference between systems based on a one-dimensional spacetime and systems based on higher dimensional spacetimes. The refinement flow in one-dimensional systems with local \footnote{Here we mean with local, actions and amplitudes which only couple variables associated to top-dimensional building blocks.} actions preserve this locality. This is (typically) not the case for higher dimensional systems, where each step of the refinement flow (typically) generates coupling between building blocks that are farther and farther apart.   

\begin{figure}[t]
\sidecaption
 \includegraphics[scale=.65]{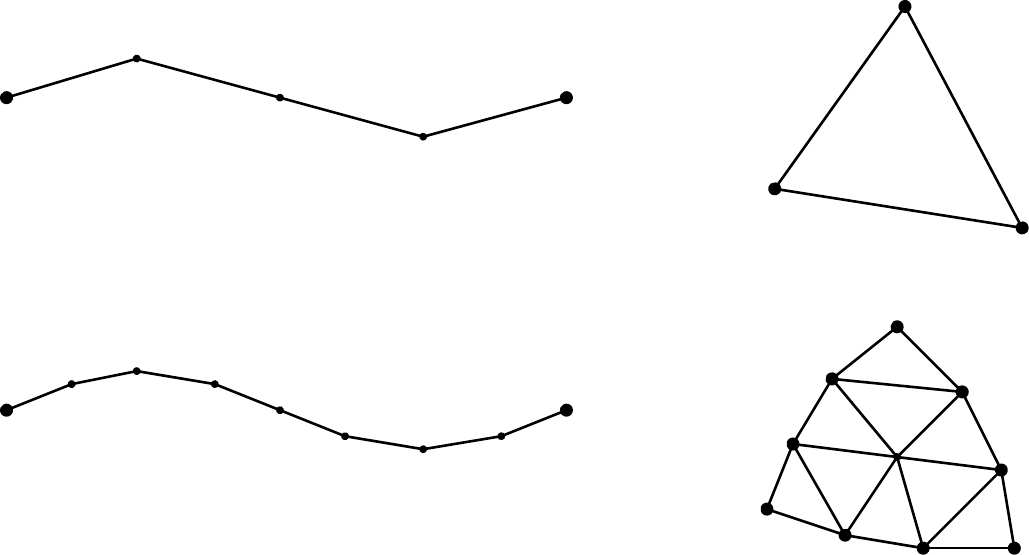}
\caption{\textit{Left:} Under refinement of a one-dimensional discrete system, the boundary data remains unchanged. \textit{Right:} For higher dimensions, refining the discretization implies a refinement of the boundary. E.g. for a discretized scalar field, a triangular building block built from nine ``smaller" triangles carries more boundary data than the smaller triangles.}
\label{fig:boundary_refinement}       
\end{figure}

The heuristic reason for this is the following: one-dimensional systems are often discretized by subdividing the one-dimensional space-time into edges. The `amount' of boundary data is the same for one edge and for a gluing of a number of edges into a `refined' edge, see Figure \ref{fig:boundary_refinement}. The finer data, which are integrated out for the refinement flow, all reside inside the refined edge. The refinement does therefore not introduce couplings between the data associated to different refined edges.

 For higher-dimensional systems, however, the amount of boundary data increases, when refining building blocks, see Figure \ref{fig:boundary_refinement}. In many refinement schemes one therefore coarse grains the boundary data of the refined blocks, that is subdivides these data into a coarse set and a fine set and integrates out all the finer data. As these finer data on the boundary are shared by at least two refined blocks, this does generically lead to couplings between these blocks. With each refinement iteration the couplings then connect  blocks that are farther and farther apart.  
 
 We will discuss in Section \ref{Sec:CBF} refinement schemes, which avoid this mechanism of producing non-local couplings. This happens by introducing a truncation for the boundary data of the refined building blocks.  Here it will be crucial to carefully choose the truncation: in fact, it should be adapted to the dynamics of the system. 
 
One example, where improved and perfect actions have been constructed is Regge gravity with cosmological constant \cite{Bahr:2009qc}.   In particular the 3D action based on homogeneously curved building blocks is perfect, as the solutions are also homogeneously curved. This provides an example where the broken diffeomorphism invariance of 3D Regge gravity with cosmological constant is regained via a refinement procedure.


Perfect actions have been also constructed for lattice QCD \cite{Bietenholz:1995cy}, and for linearized field theories with gauge symmetries, including electromagnetism and 3D and 4D gravity \cite{Bahr:2010cq, Asante:2021blx}. The choice of how to coarse grain the fields should respect the gauge symmetries \cite{Bahr:2010cq}, but there is still quite some freedom, which does influence e.g. how non-local the resulting action is \cite{Bietenholz:1997hz}. One can also import the notion of perfect action into the Hamiltonian framework and construct `perfect Hamiltonians' \cite{Lang:2017beo,Lang:2017yxi,Lang:2017oed,Lang:2017xrb}. 


Perfect amplitudes have been constructed to one-loop order for the anharmonic oscillator, for 3D Regge gravity, as well as for linearized 4D gravity \cite{Bahr:2011uj, Dittrich:2011vz,Asante:2021blx}. In fact, solving the fixed point equations derived from (\ref{Conv1}) leads to an alternative 
construction for the continuum path integral. It can in particularly be used to determine the path integral measure \cite{Dittrich:2011vz,Dittrich:2014rha,Asante:2021blx}.

Restoring reparametrization or diffeomorphism invariance for discretizations comes with a number of  further advantages:
\begin{itemize}
\item Discrete systems with diffeomorphism invariance are also discretization independent and thus perfect \cite{Bahr:2011uj,Dittrich:2012qb}. E.g. choosing discretizations with different number of building blocks leads to the same predictions. 
\item In constructing discrete actions and amplitudes one faces lots of ambiguities. As systems with diffeomorphism invariance are perfect, and thus at a fixed point of the coarse graining flow, these ambiguities are usually resolved. The work \cite{Bahr:2011uj} proofed that such ambiguities disappear for a class of one-dimensional systems.
\item  Restoring diffeomorphism invariance means that one can properly split the variables into physical observables and gauge degrees of freedom. One can also define a canonical formalism for discrete (time) systems \cite{DiBartolo:2002fu,Gambini:2002wn,DiBartolo:2004cg,Dittrich:2011ke,Dittrich:2013jaa}. If diffeomorphism invariance is not broken, one will obtain constraints, which are first class \cite{Dittrich:2009fb, Bonzom:2013tna}. This does allow to derive a consistent discretization of Hamiltonian systems with diffeomorphism symmetry. Correspondingly a (discretized) path integral with diffeomorphism symmetry, will act as a projector  onto the wave functions satisfying the constraints \cite{Bahr:2011uj,Hoehn:2014fka}. 

In contrast, for systems, where diffeomorphism symmetry is broken, one has  pseudo-constraints \cite{DiBartolo:2002fu,Gambini:2002wn,Dittrich:2009fb}. These are equations of motions, which depend very weakly on some variables --- the pseudo gauge variables. Although one could argue that the `problem of implementing the constraints' , e.g. on initial values, has disappeared, choosing the initial values freely, will typically result in solutions with unphysical properties \cite{DiBartolo:2002fu,Gambini:2002wn,Dittrich:2008pw}. 
\item
Broken diffeomorphism symmetry leads to inconsistencies for perturbation theory. These can be resolved in a framework where diffeomorphism symmetry and discretization independence are restored order by order \cite{Dittrich:2012qb}.
\end{itemize}

The first point implies  that restoring diffeomorphism invariance is equivalent to having full control over the solutions of the system. Gravity being a non-linear system, we can only hope for an approximation and truncation scheme. Often it is already helpful to obtain improved (as opposed to perfect) amplitudes. These deliver discretization independent results for sufficiently coarse observables.  We will discuss in Section \ref{Sec:CBF} a  renormalization scheme for  the construction of a family of {\it consistent} amplitudes, which $(a)$ preserves the locality of amplitudes, $(b)$ implements a refinement flow and $(c)$ implements a truncation adapted to the dynamics of the system.

\section{The notion of scale and renormalization flow of couplings in diffeomorphism invariant systems}\label{Sec:coupling}

Here we discuss the notion of scale and how the flow of coupling constants is realized in  diffeomorphism invariant systems. 
Although  coarse graining/refinement algorithms for such systems will be very similar to e.g. condensed matter systems, it is important to note the differences in interpreting the results.

For condensed matter systems one assumes initial amplitudes at a fixed small scale, referred to as lattice constant. Usually these amplitudes are seen as fundamental description, from which one derives effective amplitudes to describe the dynamics at larger scales. To this end one applies coarse graining. i.e., starting from the fundamental lattice one coarse grains the fundamental building blocks into larger blocks and correspondingly fundamental variables into coarser observables. One then integrates out finer degrees of freedom, obtaining in this way an effective dynamics for the coarser observables at a larger lattice scale. This process can be iterated, producing effective dynamics at different lattice scales. 

Often, one can describe these effective dynamics via effective actions, and these can be parameterized via coupling constants. That is, the effective dynamics at different lattice scales is described by effective actions having the same form, but with different values of the coupling constants. This defines a ``renormalization flow of the coupling constants", i.e. these are functions of the lattice scale.

For diffeomorphism invariant systems the situation is different: the initial amplitudes describe a discrete system, which can have arbitrary scale. `Scale' or the `lattice constant' are part of the basic variables. A possible measure for scale in the discretization of the reparametrization invariant system (\ref{SDRep}) is given by the differences $(t_{n+1}-t_n)$. As these are part of the dynamical variables, one needs to solve the dynamics in order to determine their values.\footnote{
The solution will generally lead to  values for the $(t_{n+1}-t_n)$, which differ for different $n$. That is, the lattice constant should be rather understood to be a function of (space-)time.
} 
 But the average size
is determined by $(t_N-t_0)/N$, and this size can be chosen  --- via the boundary data --- to be arbitrarily small or arbitrarily large. This is a general property of 
diffeomorphism invariant systems: the lattice scale is determined by the boundary data and the number of building blocks used in the discretization.

If the lattice scale is chosen to be large, the discrete actions  provide, however,  typically bad approximations to the continuum dynamics. E.g. (\ref{SDRep}) gives only a good approximation to the continuum dynamics, if $|V(q_{n+1})-V(q_{n})| \ll |q_{n+1}-q_{n}|^2 /|t_{n+1}-t_{n}|^2$. Therefore, we do not interpret a given discretization of a diffeomorphism invariant system, which breaks this symmetry, as fundamental. Rather, it serves as an approximation to the dynamics defined by the perfect action.

Different from the condensed matter case, we therefore interpret the iterative refinement procedure (\ref{CG1}) as a process of improving this approximation. Although one can extract from this refinement procedure (\ref{CG1}) a notion of flowing coupling constants, this does rather reflect the process of constructing better and better approximations to the actual dynamics, as defined by the perfect action. 

This improvement of the dynamics affects in particular the action for larger scales, that is in our example, for larger differences $(t_{n+1}-t_{n})$.  E.g. for the harmonic oscillator, a choice for an initial discretization of the 
action is given by
\ba
S_{\rm D}&=& \sum_{n=0}^{N-1}
\tfrac{1}{2} \left( 
 (q_{n+1}-q_n)^2 (t_{n+1}-t_n)^{-1} -\tfrac{\omega^2}{2}(q_{n+1}^2+q_n^2)\, (t_{n+1}-t_n) 
 \right)      \q .
\ea
The corresponding perfect action 
is given by \cite{Bahr:2011uj}
\ba \label{FPaction1}
S_{\rm Per}&=& \sum_{n=0}^{N-1}
\frac{\omega}{2} \frac{\left( \cos(\omega(t_{n+1}-t_n))(q_{n+1}^2+q_n^2) -2 q_{n+1}q_n{}
 \right)}{\sin(\omega(t_{n+1}-t_n))}      \q .
\ea
We see that $S_{\rm D}$ does approximate well $S_{\rm Per}$ for small values of $\omega(t_{n+1}-t_n) \ll 1$, but much less so for $\omega(t_{n+1}-t_n)>1$.

 The fixed point action (\ref{FPaction1}) does describe a {\it consistent} dynamics, valid over all scales. That is the dynamics at larger scales can be obtained either from probing the action directly at larger scales (as set by the boundary conditions and number of building blocks), or by probing the action at smaller scales but computing only the dynamics of variables defined at larger scales. For a perfect action these two different procedures give the same result.

Thus, the perfect or fixed point action does encode the dynamics of all scales at once, and therefore also the renormalization flow of the coupling constants. E.g. for (\ref{FPaction1}), we declared $(t_{n+1}-t_n)$ as representing a notion of lattice scale. The perfect action is a quadratic polynomial in $(q_{n+1},q_n)$. The coupling constants can be defined to be given by the corresponding coefficients, which are functions depending on $(t_{n+1}-t_n)$. These coupling constants are therefore functions of the lattice scale, given by the differences $(t_{n+1}-t_n)$. This dependence represents a physically relevant notion of renormalization flow of coupling constants.

For diffeomorphism invariant systems, such as gravity, the scale is given in terms of the spacetime metric. For reparametrization invariant field systems, one does add the spacetime coordinates as variables, and a spacetime metric can be introduced as function of these coordinates. Thus the notion of `running couplings' translates into couplings which are functionals depending on the spacetime metric.

In section \ref{Sec:need} we mentioned that one can also determine the perfect action by solving the fixed point conditions for the iterative refinement procedure (\ref{SDRep}).  These fixed point conditions constitute equations for the coupling constants as functions of the $(t_{n+1}-t_n)$, see \cite{Bahr:2011uj}. Solving these equations is analogous to solving the flow equations for the coupling constants, e.g. in the asymptotic safety approach \cite{Reuter:2019byg}. In solving the fixed point conditions one does not necessarily need to produce a solution  by determining the coupling constants for larger scales from smaller scales. In principle one can also determine the smaller scales from larger scales. As in the asymptotic safety program \cite{Reuter:2019byg}, we expect this in particular to be relevant for quantum gravity, where one rather knows the large scale dynamics, and is looking for an UV completion.

We have discussed that for reparametrization invariant systems the renormalization flow of coupling constants appears in a different guise as compared to usual lattice systems. But that  does not mean that we can avoid the need to fine tune coupling parameters in the initial discrete amplitudes. Finding the fixed points (that is perfect amplitudes) via  the refinement procedure (\ref{CG2}) does require to take infinitely many iteration steps.  Without fine tuning the infinite iterations might e.g. lead to divergent amplitudes or end up in trivial fixed points. Indeed, even for quantum mechanical systems one does need to fine tune parameters describing the path integral measure \cite{Bahr:2011uj,Bahr:2016hwc,Asante:2021blx}, in order to avoid 
reaching either a diverging  or a vanishing amplitude at the fixed point.

Although there is such a large difference in the way lattice constants and flow of coupling constants appear in systems with and without reparametrization (or diffeomorphism) invariance, the actual 
coarse graining algorithms, e.g. the tensor network algorithms discussed in Section \ref{Sec:TNR}, are often the same.

As a final note, we stress again that the  ``continuum limit'' or ``refinement limit'', as implemented by the refinement flows \eqref{CG1},\eqref{CG2} for reparametrization invarant or diffeomorphism invariant systems,  should be seen as an auxiliary construction to improve the discretization, and eventually to obtain a perfect discretization. For such systems  quantities like ``scale'' are part of the dynamical variables, we therefore do {\it not} send a lattice constant to zero. In loop quantum gravity and spin foams, observables which can be used to identify a scale, like areas and volumes, do have a discrete spectrum in the quantum theory \cite{Rovelli:1994ge}, with a minimal non-vanishing eigenvalue. This {\it physical} notion of discreteness is expected to be preserved under the refinement limit \cite{Rovelli:2011fk}.

\section{The consistent boundary formalism}\label{Sec:CBF}

Many discretized quantum systems, including spin foams, are formulated in terms of amplitudes that localize onto building blocks. The amplitude for a lattice built out of many building blocks is then given by the product over the amplitudes associated to each of the building blocks. Coarse graining such a lattice, i.e. integrating over some choice of finer data, leads generically to couplings between neighbouring building blocks. Specifically, one can expect a coupling between two building blocks to appear, if one integrates over some (finer) boundary data that are shared by both building blocks. Iterated coarse graining steps lead then to couplings between building blocks that are farther and farther apart.

Such non-local amplitudes are difficult to deal with.  They would in particular require a profound change of frameworks used in (loop) quantum gravity, e.g. the generalized boundary formalism \cite{Oeckl:2003vu,Oeckl:2005bv} or the canonical formalism. Note that a canonical formalism can also be defined for discrete systems and in particular triangulations \cite{Dittrich:2011ke}, but this requires\footnote{One can also define a canonical formalism for discrete systems on a regular lattice with non-local actions. But a framework for systems on non-regular lattices with non-local actions has, to the best knowledge of the authors, not been developed yet.} an action localized onto building blocks.

In the following we will discuss the consistent boundary formalism \cite{Dittrich:2012jq,Dittrich:2014ala}, which allows to keep the locality of amplitudes. Here we trade, in a sense, non-local couplings  for building blocks with finer and finer boundary data. Indeed, gluing building blocks (in $d\geq 2$ spacetime dimensions), each with a given amount of boundary data, we usually obtain a building block that has even more boundary data, see Figure \ref{fig:boundary_refinement}. Instead of referring to the glued building blocks as being larger\footnote{In gravity the size of the building block would be rather part of the variables.}, we say that these come with  finer boundary data.

Even if one might be able to construct the amplitudes for arbitrarily fine boundary data, one often rather wishes to compute with a certain coarse set of boundary data.  Here we wish again for a notion of  `perfect' amplitudes, i.e. amplitudes which mirror exactly those obtained in an infinite refinement limit for the boundary data.  Such a notion is given by {\it cylindrically consistent} amplitudes. 

To explain this notion, we will shortly review some formalism.  We can associate a boundary Hilbert space  ${\cal H}_{b}$ to a given building block, or more generally a spacetime region. The label $b$ will indicate the fineness of the boundary (data). 
We will assume a directed partial ordering $\prec$  on the set of labels: $b\prec b'$ means that $b'$ carries a finer set of boundary data than $b$. We will then assume that for each pair $b\prec b'$, there exists an embedding map
\ba\label{EmbM}
\iota_{b'b}: {\cal H}_{b} \rightarrow {\cal H}_{b'}     \q .
\ea
These maps have to satisfy
$
\iota_{b''b'}\circ \iota_{b'b}=\iota_{b''b}
$, for any triple $b''\succ b'\succ b$. 

The embedding maps relate coarser to finer Hilbert spaces and thus allow us to interpret coarse boundary data in terms of finer data. Given a directed partial ordered set of labels $\{b\}$, associated Hilbert spaces and consistent embedding maps, a continuum Hilbert space ${\cal H}$ can be defined as inductive (or direct) limit:
\ba\label{CHilb}
{\cal H}= \overline{ \cup_b {\cal H}_b \slash \sim} \q .
\ea
That is, one takes the disjoint union of all Hilbert space ${\cal H}_b$, but mods out by an equivalence relation. Two states $\psi_b$ and $\psi_{b'}$ are equivalent,  if there exists a refinement $b''$, with $b''\succ b$ and $b''\succ b'$, and such that $\iota_{b''b}(\psi_b)=\iota_{b''b'}(\psi_b')$.  In words, two states are equivalent, if they can be refined to the same state. To obtain a Hilbert space one takes a completion of this union. To this end, one demands\footnote{This means that the embedding maps are isometries. In quantum gravity one differentiates between kinematical and physical Hilbert spaces and inner products. One demands isometric embedding maps for the inductive limit of the kinematical Hilbert spaces. As we will discuss below, for the construction of cylindrically consistent amplitudes one should choose the embedding maps to be `dynamical'. Such dynamical embedding map might also include a projection (which implements constraints), and might therefore not be isometric. One can still demand isometry for the physical inner product, this will be however equivalent to demanding cylindrically consistent amplitudes.} that the inner products for the Hilbert spaces ${\cal H}_b$ are also cylindrically consistent, that is satisfy 
\ba
   \langle \psi_b | \chi_b\rangle_b =   \langle  \iota_{b'b}(\psi_b) | \iota_{b'b}(\chi_b) \rangle_{b'} 
\ea
for any pair $b'\succ b$. 

Canonical loop quantum gravity is based on such constructions of kinematical continuum Hilbert spaces \cite{ashtekar-lewan1,ashtekar-lewan2,thomasbook,Dittrich:2014wpa,Bahr:2015bra,Dittrich:2016typ}. Here we emphasize `kinematical', as the embedding maps implement a notion of localizable {\it kinematical} vacuum states, which can be also understood as being derived from amplitudes of topological quantum field theories \cite{Dittrich:2016typ}. The Ashtekar-Lewandowski \cite{ashtekar-lewan1,ashtekar-lewan2} vacuum is peaked on sharply vanishing electric fluxes, whereas the (possibly quantum deformed) $BF$-vacuum \cite{Dittrich:2014wpa,Bahr:2015bra,Dittrich:2016typ} is based on sharply vanishing (quantum deformed) curvature. The  embedding maps act by  putting all finer degrees of freedom into the corresponding vacuum state.  We will see below that it would be ideal to have rather embedding maps, which put the finer degrees of freedom into a {\it physical} vacuum state, that is a vacuum state defined by the dynamics of the system.

Operators ${\cal O}$ on ${\cal H}$ can be defined as  families of operators $\{{\cal O}_b\}_b$, which are cylindrically consistent, that is satisfy $\iota_{b'b}{\cal O}_{b}(\psi_b)={\cal O}_{b'}(\iota_{b'b}(\psi_b))$ for all $b'\succ b$. The different vacua and associated embedding maps do also implement different notions of coarse graining for the operators and corresponding classical phase space functions \cite{Thiemann:2000bv,Dittrich:2014wda}.

Let us assume, that we have a directed and partially ordered set of boundary Hilbert spaces $\{{\cal H}_{b}\}_b$, with consistent embedding maps, which do not necessarily correspond to the kinematical Hilbert spaces discussed above.  A {\it cylindrically consistent set of amplitudes} $\{{\cal A}_b\}_b$ is then defined to be a set of linear functionals on ${\cal H}_b$, satisfying ${\cal A}_b(\psi_b) ={\cal A}_{b'}(\iota_{b'b}(\psi_b))$ for every pair $b'\succ b$, or equivalently 
\ba\label{CCAmp}
{\cal A}_b&=& \iota^*_{b'b} \, {\cal A}_{b'} \q ,
\ea
where $\iota^*_{b'b}$ denotes the pullback of $\psi_{b'b}$.

The cylindrically consistent amplitudes allow us to formally define continuum amplitudes on the continuum Hilbert space (\ref{CHilb}). It also makes precise the notion of exactly mirroring the continuum dynamics: the embedding maps allow to interpret discrete states $\psi_{b}$ as continuum states, and the cylindrically consistent amplitudes make sure that one obtains the same results for different refinements (i.e. embeddings to different Hilbert spaces ${\cal H}_{b'}$) of a given boundary state $\psi_{b}$.

\begin{figure}[t]
\centering
 \includegraphics[scale=.325]{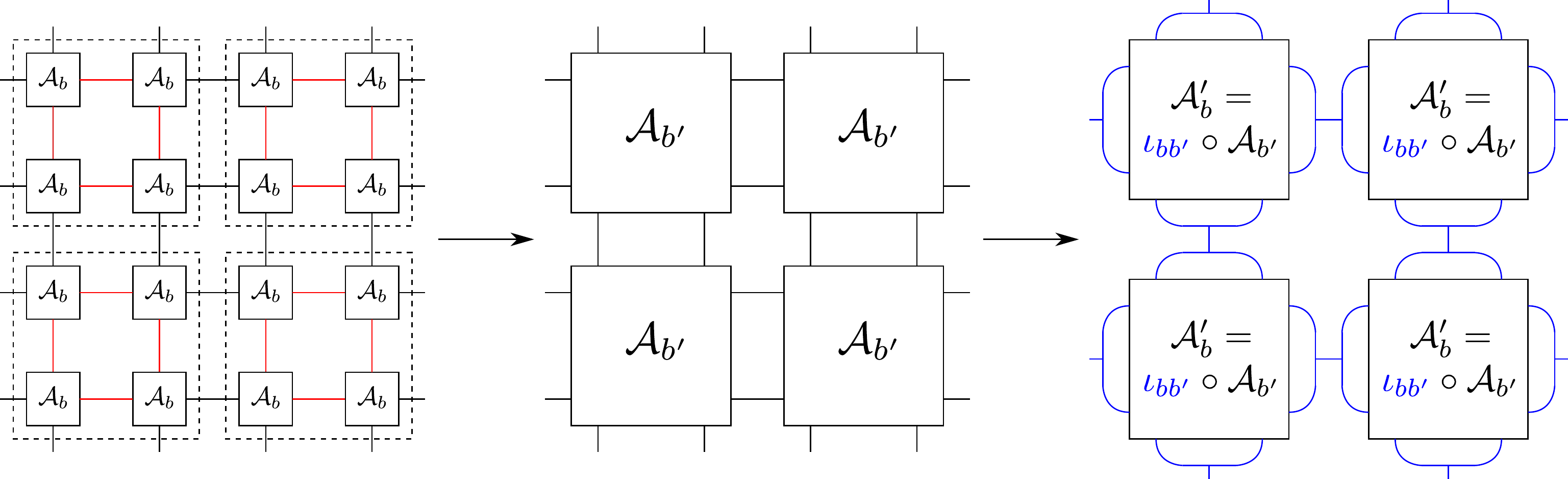}
\caption{Schematics for coarse graining spin foam amplitude: among the degrees of freedom, some (here in red) are defined as fine and summed over. The remaining effective amplitude has a refined boundary $b'$. Embedding maps $\iota_{bb'}$ in blue map to the original Hilbert space of the coarse boundary $b$ and we define the renormalized amplitude $\mathcal{A}'_b$.}
\label{fig:amplitude_renorm}       
\end{figure}

How to construct such cylindrically consistent amplitudes?  This can again be done in an iterative procedure. Starting with initial amplitudes ${\cal A}_b$ for building blocks with boundary $b$, we glue these building blocks to a block with finer boundary $b'$, see Figure \ref{fig:amplitude_renorm}. The gluing operation includes to integrate the product of amplitudes over the data that become bulk in the new building block.  In this way we obtain an amplitude ${\cal A}_{b'}$ for a finer boundary.

To improve the discrete amplitudes in an iterative scheme, similar to the ones discussed  in Section \ref{Sec:need}, we need a fixed point condition.  To this end we need to construct out of ${\cal A}_{b'}$ an amplitude for a building block with boundary $b$. Thus we use an embedding map to define an improved amplitude ${\cal A}'_b$:
\ba\label{CGFlow1}
 {\cal A}'_{b} =  \iota_{b'b}^*  \,\left(\sum_{\rm bulk}          \prod_i {\cal A}_{i,b}\right) \q .
\ea
This process can be iterated, until a fixed point is reached, that is 
\ba
 {\cal A}^{\rm fix}_{b} =  \iota_{b'b}^* \,\left( \sum_{\rm bulk}          \prod_i {\cal A}^{\rm fix}_{i,b}\right) \q .
\ea
If ${\cal A}^{\rm fix}_{b}$ does not depend on the choice of the finer boundary $b'$ in this construction, it can serve as the $b$--component of a cylindrically consistent family of amplitudes. Note that this is often not the case, if $b'$ is not sufficiently fine. But for fixed $b$ and finer and finer $b'$, one uses a finer and finer bulk and boundary discretization to compute the amplitude for a coarse boundary $b$. Thus the series of amplitudes $\{{\cal A}^{\rm fix}_{b}(b')\}_{b'}$ is expected to converge, see  \cite{Dittrich:2012jq} for examples.

The resulting fixed point amplitudes  will depend on the choice of embedding maps $\iota_{b'b}$, and might lead to non-sensible results for certain choices \cite{Dittrich:2012jq}. E.g. choosing the kinematical embedding maps described above will often result in amplitudes that just describe the associated  vacuum states of topological field theories, even if the starting point is a theory with propagating degrees of freedom.

To avoid this, it is important to determine the embedding maps from the dynamics of the system. This also means that the embedding maps might change throughout the coarse graining algorithms. Below, we will describe tensor network algorithms, which do implement embedding maps, adapted to the dynamics.

 The need for choosing embedding maps, which are informed by the dynamics, is even more important for systems with diffeomorphism symmetry. In fact, the ideal embedding maps are given by amplitudes describing the evolution of a coarser into a finer boundary. To explain this, we remind the reader that for systems with diffeomorphism symmetry the path integral does not implement a proper time evolution, but is rather a projector
 onto the space of gauge invariant, so-called physical, states. 
 
Assume we have a way to embed states associated to discrete boundaries $b\prec b'$ into a continuum Hilbert space, and that we can compute the continuum path integral `propagating' the embedded states defined from $b$ to states defined from the finer $b'$. The resulting amplitude ${\cal A}_{b'b}$ then $(i)$ project out any gauge dependence from the $b$-states, and $(ii)$ associates to all the finer degrees of freedom a physical vacuum state \cite{Dittrich:2013xwa}.
 
An ideal choice for the  embedding maps is then given by these refining amplitudes $\iota_{b'b}={\cal A}_{b'b}$. The reason is that with this choice, we satisfy the cylindrical consistency condition (\ref{CCAmp}) for the amplitudes ${\cal A}_b\equiv {\cal A}_{b\emptyset}$ \cite{Dittrich:2014ala}. Here $ {\cal A}_{b\emptyset}$ is constructed as described above, with $b$ being a general boundary label, and $\emptyset$ denoting the `empty' boundary with no boundary data. That is we compute the continuum path integral for a building block with one boundary $b$, which can be interpreted as no-boundary state \cite{Oeckl:2003vu,Dittrich:2011ke}.

The consistency condition can now be expressed as
\ba\label{CCAmp2}
{\cal A}_{b \emptyset } &=&    {\cal A}_{bb'} \circ   {\cal A}_{b'\emptyset}  \q ,
\ea
where $\circ$ includes integrating or summing over the data associated to the boundary $b'$. This has to hold for any $b' \succ b$.
We assumed that the amplitudes result from a continuum path integral, they should therefore be discretization independent. Eq. (\ref{CCAmp2}) does express such a notion of discretization independence: namely that the final amplitude should not depend on the number of building blocks or evolution steps. 

We thus see that for diffeomorphism invariant systems, the embedding maps should be given by the amplitudes. In this way the embedding maps do also describe the physical vacuum state for the given system.  These concepts can be realized for topological quantum field theories \cite{Dittrich:2013xwa,Dittrich:2016typ}.
 
Naturally, the situation is much more involved for systems with propagating degrees of freedom. Often, one can only resort to numerical coarse graining algorithms. We will discuss next tensor network algorithms, which implement a version of the coarse graining flow (\ref{CGFlow1}), with a dynamically determined embedding map. In fact, embedding maps are constructed using a localized form of the transfer operator method, which also serves to identify vacuum states.


\section{Tensor Network Renormalization }\label{Sec:TNR}

A concrete numerical realization of the consistent boundary formulation for physical systems are tensor network methods. Here we mostly discuss the ``Tensor Network Renormalization Group'' (TNRG) \cite{Levin} as variations of it were used in all spin foam related research. Similar ideas were originally invented for 1D systems, called ``Density Matrix Renormalization Group'' \cite{White:1992zz}, and TNRG can also be understood as a generalization of corner transfer matrix methods for 2D systems \cite{Orus:2011nj}. Tensor network coarse graining methods were further developed to include entanglement filtering \cite{VidalEntFiltering,mera}, with ``Tensor Network Renormalization'' as its most recent variant \cite{GuWen,vidal-TNR,Hauru:2017jbf}. 

The key idea of TNRG is to cast the partition function of a physical system into the form of a tensor network, a contraction of tensors, and then manipulating this network to approximate the partition function by a coarser network of effective tensors. This defines a renormalization group flow of tensors, while keeping the description of the system \textit{local}: tensors only interact with their direct neighbours. Here, a tensor is a multidimensional array associated to a region of the physical system. Its labels correspond to (a basis of) the boundary data of that region, and for each choice of boundary data it stores the associated amplitude. In this sense, a tensor is similar to spin foam amplitude(s) as it associates an amplitude to a region as a function of its boundary data \cite{Dittrich:2011zh,Dittrich:2014mxa}. We represent such a tensor pictorially by a node with several legs, one leg for each index of the tensor, which in turn corresponds to a boundary datum. Note that these tensors do not make a reference to a background spacetime; they merely encode the degrees of freedom of the system. In this regard they are well suited to be applied to background independent approaches to quantum gravity.

Having identified a tensor as the amplitude associated to a region of the system, the next step is to glue tensors together. To do so we identify the shared boundary data of two tensors and sum over all possible values this shared data can take, i.e. we contract their indices. 
Pictorially we represent this contraction by connecting the respective legs of the tensors. Continuing this process, we build a network of connected tensors, whose contraction defines the partition function.

The goal of a coarse graining step is to approximate the original tensor network by a coarse tensor network by renormalizing the tensors. The coarse tensor network is then the starting point for the next iteration of the algorithm. There exist many realizations of the TNRG and they all have two steps in common: explicit summation over fine (bulk) degrees of freedom as well as variable transformations and truncations of (boundary) degrees of freedom to define effective boundary Hilbert spaces. These variable transformations play the role of (inverse) embedding maps mentioned above, as they directly translate fine into coarse degrees of freedom. Crucially, the embedding maps are computed from the tensors in each iteration of the algorithm via a singular value decomposition. Thus, the embedding maps can change in each iteration and are in fact chosen according to the dynamics encoded in the tensors.

Without going into the details of the algorithms, we want to briefly explain the role and importance of the singular value decomposition (SVD) in defining the embedding maps. A SVD decomposes a matrix $M$ into two unitary matrices $U$ and $V$ and a diagonal matrix $\lambda$, $M = U \lambda V^\dagger$. It is applicable to real and complex matrices, where $M$ need not be a rectangular matrix. The matrices $U$ and $V$ are matrices of left and right singular vectors respectively, the entries $\lambda_i$ of the diagonal matrix $\lambda$ are called singular values. Crucially, the singular values are real, non-negative and ordered in size, $\lambda_1 \geq \lambda_2 \geq \dots \geq \lambda_N \geq 0$. Hence, the size of the singular values relative to the largest one determine how relevant the associated singular vectors of $U$ and $V$ are to reconstruct $M$. On the other hand, we can define the optimal approximation (in the least square sense) of $M$ by a matrix of rank $D < N$ by setting all singular values $i > D$ to zero \cite{Levin,RevModPhys.86.647}. So, if the size of singular values decreases rapidly relative to the largest ones, it is possible to well approximate the matrix $M$ by a matrix of lower rank.

\begin{figure}
\centering
\includegraphics[scale=.525]{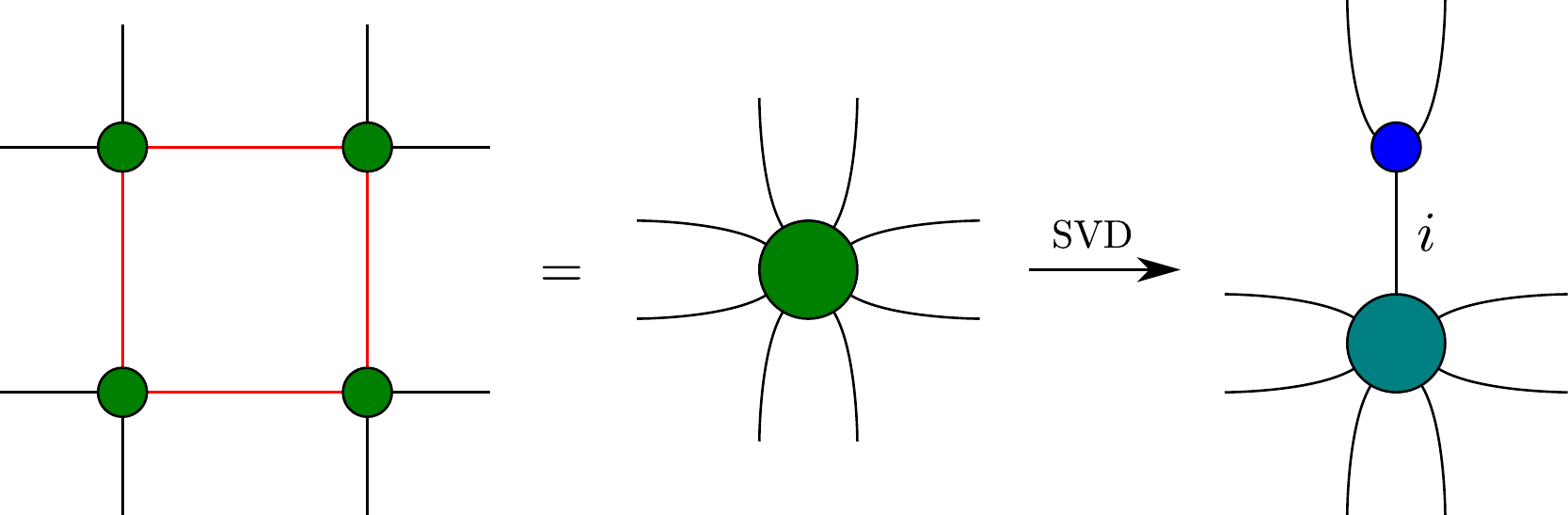}
\caption{After the contraction of fine indices (red), we obtain a new tensor with eight legs. To decompose it, we turn it into a matrix by grouping its indices into two groups: the upwards pointing indices form one group, with all remaining indices in the other. After the SVD, we obtain two tensors, a 3-valent and a 7-valent one. We use the 3-valent one as an embedding map, as it translates fine into coarse degrees of freedom.}
\label{fig:tensor_splitting}       
\end{figure}

How can this matrix decomposition be used to coarse grain tensors or amplitudes? Ideally we would like to directly split tensors into multiple tensors of lower rank. We can then recombine those into new effective tensors by contracting the original indices. Unfortunately, a decomposition analogous to SVDs does not exist for tensors. Instead, we rearrange the tensor into a matrix, by grouping its indices into two matrix indices. These matrix indices simply label the product basis of the boundary Hilbert spaces of the grouped tensor labels. Now, we decompose this matrix using a SVD: the matrices $U$ and $V$ now define the variable transformations of the old product basis into a new effective one labeled and ordered by the singular values. We show an example of such a decomposition in fig. \ref{fig:tensor_splitting}. From the relative size of these singular values then directly follows the relative relevance of the corresponding singular vectors for reconstructing the tensor. With this information, we truncate the most irrelevant singular values / vectors, e.g. by always truncating after a fixed number of singular values or only considering singular values above a certain size relative to the largest one. Once we have obtained the singular vectors, we translate the product basis back into the original tensor indices and thus define a split tensor. 

\begin{figure}
\sidecaption
\centering
\includegraphics[scale=.45]{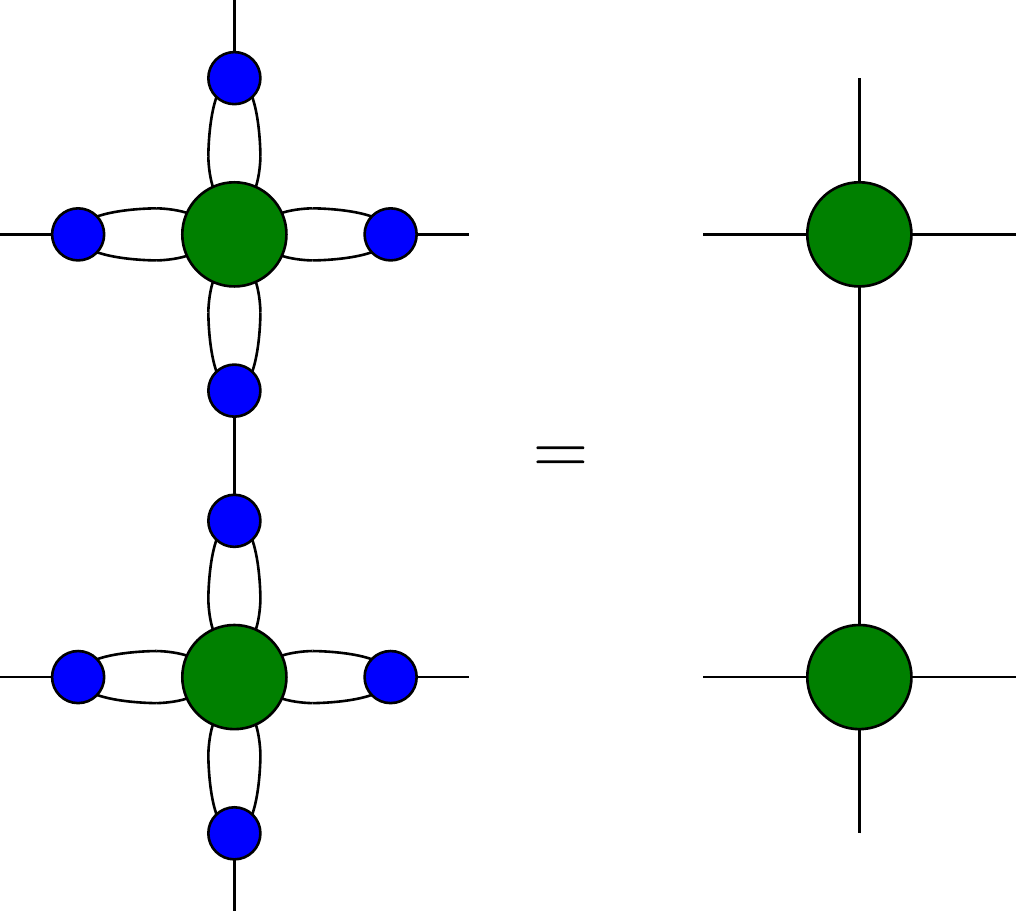}
\caption{After the contraction and computation of embedding maps via a singular value decomposition, we contract all fine boundary legs of the effective tensor with embedding maps. They translate fine into coarse degrees of freedom and introduce truncations. A coarse network of renormalized tensors remains.}
\label{fig:TNW_algo}       
\end{figure}

One step of a coarse graining algorithm is illustrated in figs. \ref{fig:tensor_splitting} and \ref{fig:TNW_algo}. First, we sum over fine degrees of freedom resulting in a new tensor with more boundary data. We rearrange its indices with the goal to split into two tensors, a 3-valent one containing the two upwards-pointing indices and a 7-valent one with the remaining indices. The new edge of the tensors will be labeled by the singular values obtained in the decomposition. The 3-valent tensor defines a unitary embedding map.
Without truncations they act as isometries between different basis of the boundary Hilbert spaces. However, truncations are necessary, since the coarse boundary Hilbert spaces grow exponentially with each iteration. Keeping track of the size of the truncated singular values (relative to the largest ones) makes it possible to characterize the truncation error. In the final step, the fine boundary data of the tensors get contracted with the derived 3-valent embedding maps, resulting in a renormalized tensor with coarse boundary data.

\subsection{Tensor networks and spin foams}

In many ways, TNRG is well suited to be applied to spin foams. The method does not require a background spacetime, the concept of tensors and their boundary data is similar to spin foam amplitudes and the method, in contrast to Monte Carlo sampling, works well for complex and oscillating amplitudes 
: in TNRG we explicitly sum over fine degrees of freedom instead of sampling over them. Additionally, TNRG coarse graining is performed locally, it is not necessary to consider the entire system at once. However, spin foams cannot be readily cast into tensor network form for two reasons: firstly, tensor networks are built under the assumption that degrees of freedom are only shared between two tensors. For spin foams this condition is not fulfilled:
 here, tetrahedra and triangles are decorated by algebraic data. While a tetrahedron is only shared among two 4-simplices, a triangle can belong to several 4-simplices. Assigning a tensor to each 4-simplex, only a part of the spin foam state-sum can be expressed as a contraction of a tensor network.
This insight is useful for optimizing numerical algorithms \cite{Gozzini:2021kbt}, 
yet it is not sufficient for coarse graining \`a la TNRG. As we will argue below, spin foams require a generalization of this formalism called decorated tensor networks.

The second obstacle is that TNRG methods require finite tensors, e.g. with finite dimensional boundary Hilbert spaces. In spin foams the representation labels on faces are typically unbounded. Introducing a cut-off on representation labels breaks gauge symmetry. A notable and physically relevant exception are spin foams with non-vanishing cosmological constant, where the gauge-invariant cut-off is inversely proportional to the square-root of the cosmological constant. In recent years, several different spin foam models have been proposed \cite{Han:2010pz, Fairbairn:2010cp, Haggard:2015ima, Haggard:2015yda, Han:2021tzw}. One well-known 3D spin foam model with cosmological constant is the Turaev-Viro model \cite{Turaev:1992hq}, which uses a quantum deformation of SU$(2)$ \cite{biedenharn}. Such quantum deformations are also used in TNRG applications to (analogue) spin foam models \cite{Dittrich:2013voa, Dittrich:2013aia, Dittrich:2016tys}.

\subsection{Models with global symmetry: spin net models}

TNRG methods are primarily aimed at condensed matter systems with a global symmetry, the 2d Ising model being a prime example \cite{Levin, vidal-TNR}. Without external magnetic field it possesses a global $\mathbb{Z}_2$ symmetry. Thanks to this property, it is straightforward to cast this model into tensor network language. One way to do so is to expand the system in terms of irreducible representations of its symmetry group, analogous to the familiar harmonic analysis of SU$(2)$ used in LQG and spin foams. The boundary Hilbert spaces are tensor products of representation spaces of the underlying group where the tensor encodes the coupling rules. This construction (and the differences and similarities to spin foams) was first presented in \cite{Bahr:2011yc} for finite groups before it was extended to quantum groups in \cite{Dittrich:2013aia, Dittrich:2013voa}. Such models with global symmetries are called spin net models.

TNRG methods were applied to 2D spin net models for various symmetry groups, namely finite Abelian groups $\mathbb{Z}_q$ \cite{Dittrich:2011zh}, the finite non-Abelian permutation group $S_3$ \cite{Dittrich:2013uqe} and the quantum group SU$(2)_k$ \cite{Dittrich:2013voa}. Models with SU$(2)_k \times$ SU$(2)_k$ symmetry \cite{Dittrich:2016tys} mimic the construction of Euclidean 4D spin foam models, such as the Barrett-Crane \cite{Barrett:1997gw} and the EPRL model \cite{Engle:2007wy}. These papers show a gradual increase in complexity of the models, e.g. non-Abelian models are more challenging than Abelian ones due to higher dimensional representation spaces. Additionally, representation theory is used to optimize the algorithm and improve the interpretation of the coarse graining flow. Using the coupling rules, we can bring the matrix that we intend to split by a SVD into a block diagonal form, where each block is labeled by a representation label. Then, we perform a SVD for each block, which is faster due to the smaller size of the matrices, and label the new tensor indices with the block label. Thus, we maintain the original Hilbert space interpretation of the model, allowing for a more physical interpretation of the flow and fixed points. For more detailed explanations we refer the interested reader to the respective papers \cite{Dittrich:2011zh, Dittrich:2013uqe, Dittrich:2013voa, Dittrich:2016tys} or this recent review \cite{Steinhaus:2020lgb}.

The TNRG method is ideal to identify different phases of the model and map out its phase diagram. Here we define a phase as the region in parameter space for which the system flows to the same attractive fixed point under renormalization. We distinguish these fixed points by how many non-vanishing singular values are associated with the representation labels. Note that all these fixed points possess only finitely many degrees of freedom and thus describe topological theories.

In case the theory possesses multiple phases, we find phase transitions, which can feature interesting dynamics. Using TNRG methods, we gain insights into whether a phase transition is of first or higher order. For a first order transition the system quickly converges to either of the final two fixed points, no matter how close we tune the parameters to the transition. The behavior is starkly different for transitions of higher order: close to the phase transition the system remains almost the same for a number of iterations before it converges to either of the attractive fixed points. Moreover, the closer one tunes towards the transition, the system remains almost the same for more and more iterations. This indicates the presence of a repulsive fixed point and an (almost) scale invariance. Additionally, close to this transition, the singular values only slowly decrease in size and truncations cannot be justified as the largest truncated singular value is close to the smallest kept one. This shows that more and more degrees of freedom are relevant, which hints at a diverging correlation length. In tensor network language, where we describe the system by locally interacting amplitudes, this translates to infinitely large boundary Hilbert spaces.

Before we discuss renormalizing gauge theories (for finite and quantum groups) let us briefly mention some disadvantages of TNRG that were addressed in later schemes like Tensor Network Renormalization (TNR) \cite{vidal-TNR}. In particular for higher accuracy, i.e. taking many singular values into account, TNRG converges to ``unphysical'' fixed points. They are of a particular form 
and highly dependent on the initial conditions, which can obstruct identifying the correct renormalization group flow and universality classes. This issue is addressed by entanglement filtering: it identifies and removes short scale entanglement, which avoids the occurrence of these particular fixed points, and leads to a proper renormalization group flow and a better approximation of fixed points on second order phase transitions.

\subsection{Local 
symmetries: decorated tensor networks and the fusion basis}

As mentioned above, for systems with local gauge symmetry, like spin foams, we cannot readily apply methods like TNRG. 
Gauge systems encode redundantly degrees of freedom into local variables, which are subject to Gauss constraints. This increases the computational costs for TNRG, as it scales exponentially with the number of variables per building block. 

\begin{figure}
\sidecaption
\includegraphics[scale=.35]{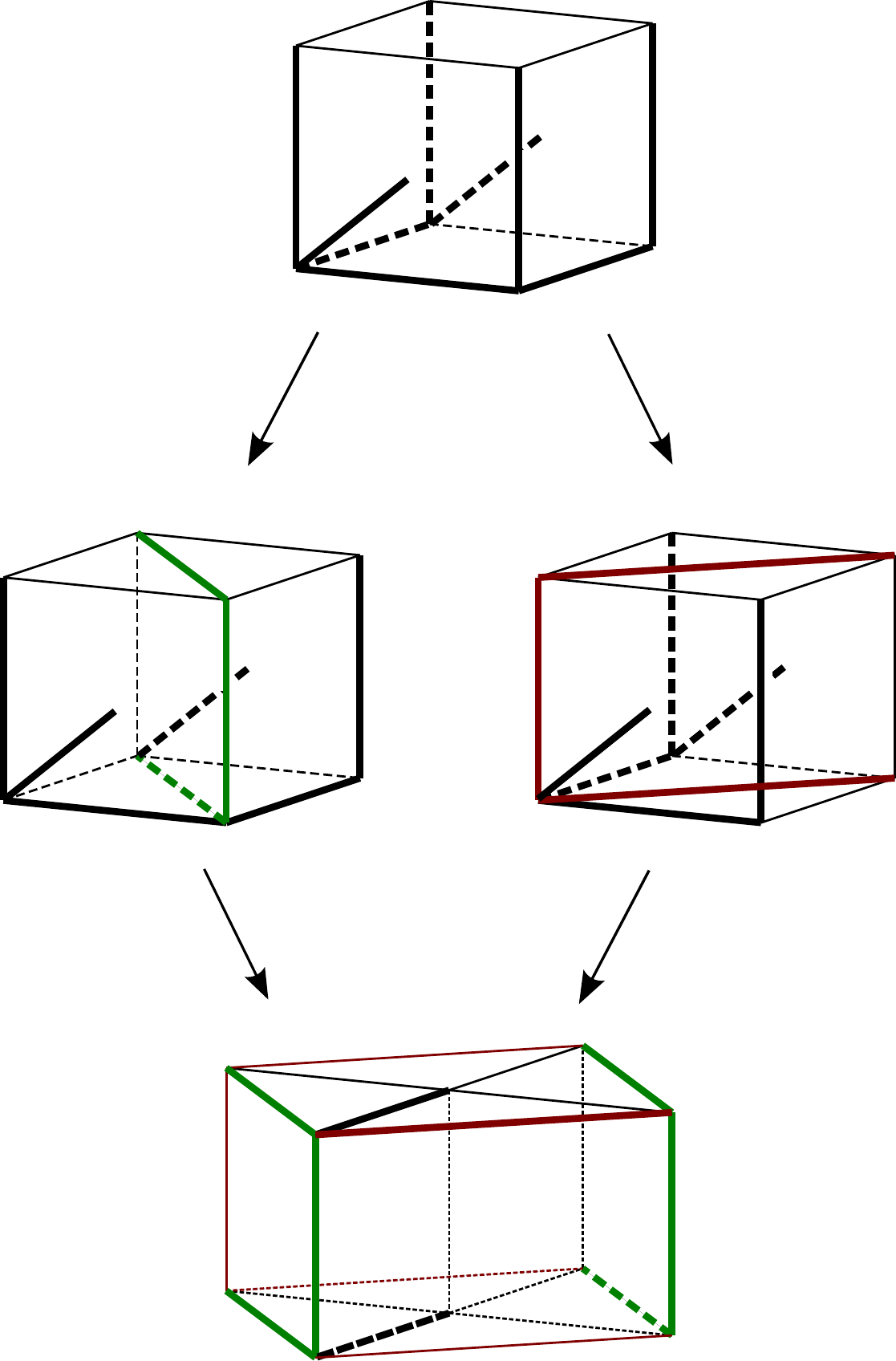}
\caption{Graphical illustration of the decorated tensor network coarse graining algorithm for Abelian lattice gauge theories: The boundary Hilbert space is spanned by spin network states. The bold edges show the non-gauge fixed variables. These cubes are again split by a singular value decomposition in two different ways. The variables must be adapted to the splitting, e.g. we must have data shared between both split amplitudes. Eventually, the resulting prism amplitudes are recombined into a coarse grained cube.}
\label{fig:3d_splitting}       
\end{figure}

Instead of expressing lattice gauge theories strictly in tensor network language, \cite{Dittrich:2014mxa} proposes to use decorated tensor networks. 
Here decorations refer to labels associated to lower dimensional boundary components. We consider the amplitudes associated to spacetime regions and their boundary states, e.g. 3D cubes equipped with spin network states, as the fundamental objects of the system which we intend to renormalize. Coarse graining is performed analogous to TNRG methods: these amplitudes are split by using a SVD and then rearranged into new effective ones, where we sum over bulk degrees of freedom. The key to this algorithm is the boundary Hilbert space: firstly, to implement this procedure economically, we solve for all Gauss constraints and explicitly express redundant variables as functions of the remaining ones. This corresponds to choosing a maximal tree in the spin network. Secondly, this maximal tree must be chosen with the intended splitting in mind: the remaining variables are distributed equally across the two split amplitudes and the data shared between them. Given this distinction, we define matrices that can be split by a SVD. Variables belonging to either split amplitudes are summarized in either index of the matrix, while we keep the shared variables fixed. We then perform the decomposition for each shared configuration, which reduces numerical costs and ensures the explicit reconstruction of the amplitude in terms of the original degrees of freedom. Finally, the maximal tree is adapted again for the gluing of partial amplitudes into a final effective one. This algorithm is illustrated in fig. \ref{fig:3d_splitting}. 
To lowest accuracy, we keep a single singular value per shared configuration of the original amplitude. In case we retain more singular values, each face of the renormalized amplitude carries a tensor network of additional indices with one index per face. This gives the name to this algorithm as the tensor network is ``decorated'' with an intricate boundary Hilbert space.
This algorithm was successfully applied to 3d $\mathbb{Z}_2$ lattice gauge theory and identified the phase transition between weak and strong coupling phase to reasonable accuracy \cite{Dittrich:2014mxa}.

Building on this foundation, the algorithm was extended to 
non-Abelian lattice gauge theory in \cite{Delcamp:2016dqo} and tested successfuly for the finite group $S_3$. The general procedure is similar to the Abelian case, however the basis transformations are considerably more complicated as they involve a switch between spin network and holonomy basis via group Fourier transform \cite{Bahr:2011yc}. 



Beyond this successful extension of the algorithm to non-Abelian theories, it also underlines two important facts: firstly, the choice of variables, and thus observables described, is crucial in coarse graining algorithms. To preserve the interpretation of the theory, the renormalized theory must be capable of capturing the coarse observables associated with its underlying discretization. To do so, we must choose coarse observables and variables among the original fine ones and adapt the algorithm accordingly to describe these observabes. Secondly, for non-Abelian theories, the spin network basis is unstable under coarse graining, as Gauss constraint violations, also known as torsion degrees of freedom, appear from originally gauge invariant amplitudes \cite{Livine:2013gna,Dittrich:2014wda,Livine:2017xww}.

Both aspects are solved by using the fusion basis for 3D lattice gauge theory. The fusion basis \cite{Dittrich:2016typ, Delcamp:2016yix} is a gauge-invariant basis for lattice gauge theories that a) organizes the degrees of freedom according to a coarse graining (also known as fusion) scheme and b) is able to represent torsion degrees of freedom. 

For example, consider a 3D cube with six plaquettes. Each plaquette can carry a curvature and torsion excitation. This excitation can be measured by a (generalized) Wilson loop around a given plaquette and is characterized by a pair of quantum numbers. Pairs of plaquettes can be fused --- correspondingly one can fuse the elementary plaquette 
excitations. These fused excitations can be again characterized by a pair of quantum numbers, arising from a Wilson loop around the fused plaquettes, and so on. This information can be encoded into a (fusion) tree, which has one leave for every elementary plaquette, and carries a pair of quantum numbers on every leave and branch. The quantum numbers are subject to fusion rules, whose precise form depends on the gauge symmetry (quantum) group.

\begin{figure}
\centering
\includegraphics[scale=.525]{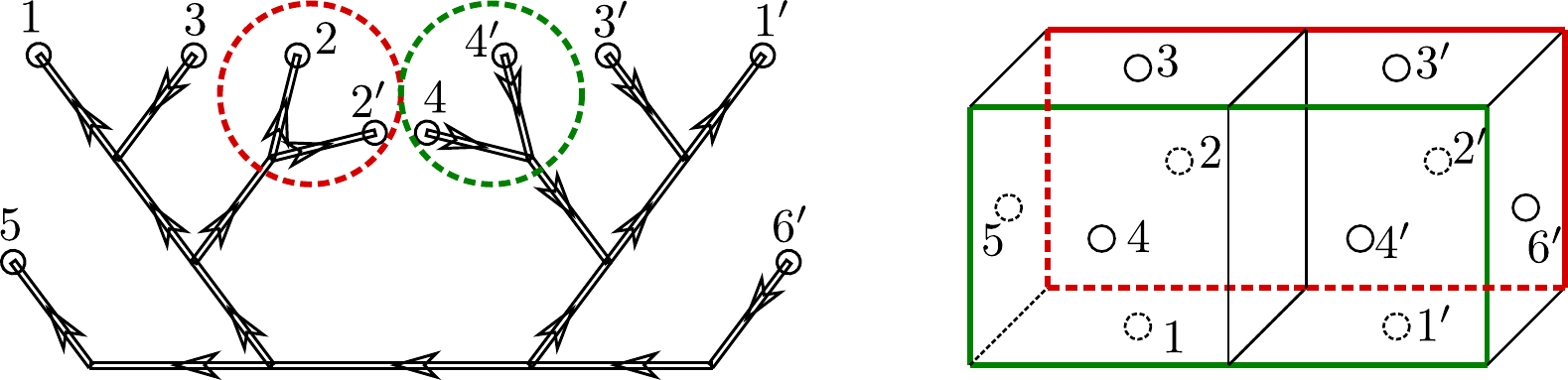}
\caption{Choice of fusion tree for coarse graining. The left side shows a fusion tree for two glued 3d cubes. Each end represents an elementary plaquette. The shown fusion tree directly fuses pairs of elementary plaquettes (green and red) into coarse plaquettes.}
\label{fig:fusion_tree}       
\end{figure}

Gluing cubes to larger blocks we also glue the trees to larger trees. We can then apply a tree (i.e. variable) transformation, such that we can fuse the excitations associated to elementary plaquettes, which are coarse grained to effective plaquettes for the larger building blocks, see Fig. \ref{fig:fusion_tree}. Note that this allows us access to the (coarser) Wilson loop observables around these effective plaquettes.


\cite{Cunningham:2020uco} details the construction of a decorated tensor network algorithm for 3d lattice gauge theories with underlying quantum group SU$(2)_k$ symmetry. As outlined the algorithm allows access to coarse grained Wilson loops. One application is a test 
of the area law for Wilson loop operators in the strong coupling regime \cite{Cunningham:2020uco}. The algorithm allows us to identify strong and a weak coupling regimes and the study of how the critical coupling changes as function of the quantum group level $k$ \cite{Cunningham:2020uco}.


In a nutshell, TNRG based coarse graining algorithms are a concrete numerical realization of the consistent boundary formulation implemented and tested for 2D spin net models and 3D models with local Abelian and (q-deformed) non-Abelian gauge symmetry. Embedding maps are derived directly from the amplitudes and define effective coarse degrees of freedom from fine ones depending on their relevance to the dynamics. Strong arguments for TNRG methods are their robust ability to identify different phases of the theory, exploration of the system near phase transitions and their applicability to models with (complex) oscillatory amplitudes. Models with local gauge symmetries require special care, due to redundant local variables as well as the instability of the spin network basis under coarse graining. The fusion basis in 3d is an intriguing solution to these problems, as it is stable under coarse graining and allows us access to coarse (Wilson loop) observables.

Beyond the development of (decorated) tensor network methods for 4D spin foams and lattice gauge theories, spin foams pose additional challenges. Efficient algorithms for evaluating spin foam amplitudes and summing over (fine) degrees of freedom are indispensable. Thanks to progress over the last few years \cite{Dona:2017dvf, Dona:2019dkf, Gozzini:2021kbt, Asante:2020qpa, Asante:2020iwm, Asante:2021phx,Han:2020npv, Han:2021kll}, it is hopefully only a question of time until coarse graining of full 4D spin foams can be tackled. So far, such explorations were only possible by significantly restricting spin foam configurations, which we will detail below.

\section{Restricted spin foam models}\label{Sec:Restr}

\begin{figure}
\sidecaption
\includegraphics[scale=.4]{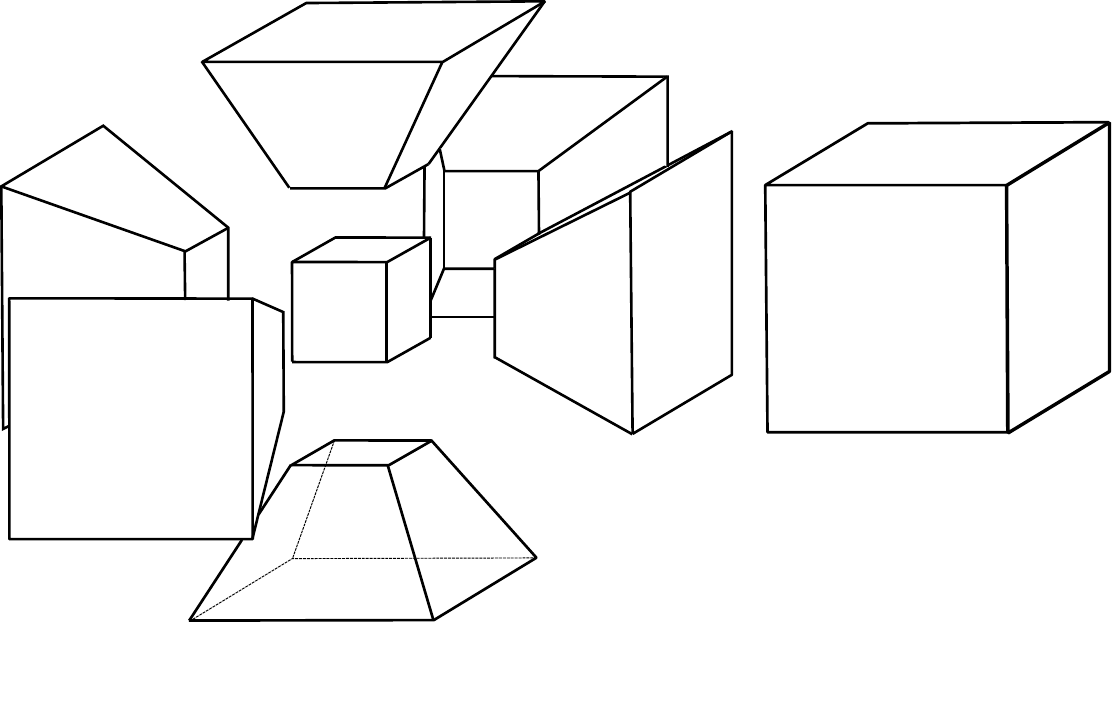}
\caption{The 3d boundary of a hyperfrustum: Two 3D cubes of different size are connected to each other via six 3D frusta, generalizations of trapezoids.}
\label{fig:frusta}       
\end{figure}

The main idea of restricted spin foam models is to define a subset of the full spin foam path integral. To develop an iterable coarse graining method, it is convenient to choose a 2-complex more general than a triangulation, e.g. one with hypercubic combinatorics. An extension of the Riemannian EPRL model \cite{Engle:2007wy} to general 2-complexes was defined 
in \cite{Kaminski:2009fm}. Instead of summing over all possible shapes of polyhedra, encoded as intertwiners, only specific ones are allowed, typically given by coherent Livine-Speziale intertwiners \cite{Livine:2007vk} peaked on the shape of a classical polyhedron, e.g. a cuboid. Depending on the polyhedron at hand, additional restrictions on the areas of the faces might arise, which furthermore reduce the number of summations in the path integral. In this case, polyhedral restrictions were proposed in \cite{Bahr:2015gxa, Bahr:2017eyi, Bahr:2018ewi, Assanioussi:2020fml}. Here we will focus on cuboids \cite{Bahr:2015gxa} and frusta \cite{Bahr:2017eyi, Bahr:2018ewi} as these were investigated in the context of renormalization. Cuboids are the 3D analogue of rectangles, where all angles between faces are rectangular and the areas of opposite faces must agree. Eight cuboids are then combined into a hypercuboid, which describes flat space-time with metric discontinuities and torsion \cite{Bahr:2015gxa}. Frusta on the other hand are 3D generalizations of trapezoids and can be seen as a generalization of the cuboid case. In 4D, two cubes and six frusta form a hyperfrustum, which describes the transition from a cube to a cube of a different size. In fig. \ref{fig:frusta} we show illustrate the 3D boundary of a hyperfrustum. Therefore this model can be understood as a potential cosmological subsector of spin foam models describing the expansion / contraction of 3D cubulations \cite{Bahr:2017eyi}.

The second vital ingredient is the semi-classical approximation of the spin foam amplitudes, in particular the vertex amplitudes. These are hard to compute in the quantum regime, even more so for hypercubic 2-complexes \cite{Allen:2022unb}: the vertex amplitude is defined as a contraction of more intertwiners compared to a 4-simplex and these intertwiners are of a higher valence. The increase in numerical costs is drastic, which makes it impossible to explore the amplitude beyond small spins. While further numerical optimizations are possible, approximating the full vertex amplitude by its semi-classical approximation is indispensable to study coarse graining. The derivation of a semi-classical formula via stationary phase analysis is completely analogous to the 4-simplex case \cite{Conrady:2008mk,Barrett:2009gg}. Thanks to using fixed coherent intertwiners, it is straightforward to derive a closed formula for the asymptotic expression of the vertex amplitude. To summarize, intertwiner restrictions and the asymptotic formula for spin foam amplitudes provide readily computable amplitudes as well as restrictions on the configurations to sum over. These drastic simplifications make spin foam model simulations on large 2-complexes with multiple vertices accessible.

After this basic outline on the definition of the model, let us briefly explain the coarse graining method as it uses the ideas of the consistent boundary formulation, yet employs further approximations to derive a renormalization group flow. Again, embedding maps are used to relate boundary Hilbert spaces, e.g. the one of a subdivided hypercuboid to the Hilbert space of a coarse hypercuboid. These embedding maps are, however, not dynamical, and instead geometrically motivated. Take again the example of a hypercuboid: a refined hypercuboid, consisting of 16 hypercuboids, gives rise to a coarse hypercuboid, whose square areas are given by the total area of the subdivided squares. Hence, a coarse hypercuboid amplitude is defined as a superposition of all refined amplitudes giving rise to this coarse geometry.


In contrast to TNRG methods, the amplitude itself is not used to derive the renormalization group flow; instead we compare expectation values of observables for a coarse and fine 2-complex while keeping the boundary data fixed, such that we represent the same transition in both cases. These expectation values are computed by numerical integration techniques \cite{Hahn:2004fe}. For the results to agree and the theory to give consistent predictions, we must adapt the parameters in the coarse and fine cases. If the expectation values (approximately) match, the coarse amplitude can be understood as the effective description of the fine one, which defines a renormalization group flow of the amplitudes projected onto the original parametrization of the amplitude.
Because of this projection, this method washes out details of the renormalized amplitude. Therefore, as for other renormalization methods, we obtain a flow under certain assumptions and truncations, which necessarily must to be eventually lifted to check its viability.

For the renormalization group flow studied for cuboids \cite{Bahr:2016hwc,Bahr:2017klw} and frusta \cite{Bahr:2018gwf}, one parameter plays a crucial role, the exponent of the face amplitude. It is added by hand to reflect an ambiguity in the definition of the EPRL model\footnote{This ambiguity can be fixed by demanding invariance of the face amplitude under subdivisions of the faces \cite{Bianchi:2010fj}.} without changing the overall form of the amplitude, and can be seen as a modification of the path integral measure. In the cuboid case, the face amplitude can be translated into (a certain power of) the volume of a hypercuboid. A similar path integral measure has been discussed in quantum Regge calculus \cite{Hamber1999}. This exponent has a strong effect on the scaling behavior of the amplitude affecting in the renormalization group flow: for both cuboids and frusta indications for a UV-attractive fixed point were found, which indicates a second order phase transition. Indeed, this might be related to a restoration of (an Abelian subgroup of) diffeomorphisms. Furthermore, in the frusta setting the parameter space included gravitational and cosmological constants, which might be related to UV-repulsive directions (for the transitions considered). Further research, in particular lifting of the assumptions mentioned above, is necessary to show whether these results are robust.

In addition to coarse graining calculations, the restricted spin foam models additionally opened the door towards exploring observables on such quantum space-times, e.g. the spectral dimension \cite{Steinhaus:2018aav} and spin foam matter systems \cite{Ali:2022vhn}. However, while these first insights are valuable, the restrictions on the spin foam path integral must eventually be lifted. Unfortunately, the large numerical costs in the quantum regime \cite{Allen:2022unb} are an obstacle, which is less pronounced for triangulations. An interesting approach, which picks up the idea to utilize the semi-classical approximation of the vertex amplitude while barely restricting the path integral are effective spin foam models.

\section{Effective spin foam models and their refinement limit}
\label{Sec:ESFs}

The various methods for determining the renormalization flow via iterative coarse graining and refining require some control on the dynamics of the underlying discrete theory. Spin foam models on the other hand have complex and complicated amplitudes, making the assessment of spin foam dynamics difficult even for quite coarse triangulations. Although there have been recent improvements of numerical techniques for spin foams \cite{Han:2020npv,Gozzini:2021kbt,Dona:2017dvf,Dona:2019dkf,Dona:2018nev}, the amount of numerical resources required remains challenging. 


Recently, a new family of gravitational spin foam models, which go by the name `effective spin foams', have been constructed in  \cite{Asante:2020qpa,Asante:2020iwm,Asante:2021zzh}. These new models have numerically accessible amplitudes with a transparent structure, 
while keeping the universal dynamical properties of spin foams.  This has allowed unprecedented computations of expectation values \cite{Asante:2020iwm} and surprising insights into the dynamics of the refinement limit \cite{Dittrich:2022yoo}.  
In what follows, we discuss effective spin foams and  their (perturbative) refinement limit. 

Like all spin foam models, effective spin foams are path integrals over quantum geometries, derived from loop quantum gravity, and associated to triangulations. Such geometries can be labelled by the discrete eigenvalues for the areas in the triangulation. One has furthermore so-called intertwiner labels encoding the quantum shape of the tetrahedra \cite{Baez:1999tk}. In effective spin foams these are assumed to be integrated out for the bulk tetrahedra. This does  reduce the number of degrees of freedom considerably and provides one reason for the numerical efficiency of the models.

Spin foam construction is based on an extended configuration space, which is shared by a topological field theory. The latter fact allows an exact quantization of this extended configuration space \cite{Perez:2012wv}. To obtain gravity one has to implement simplicity constraints, which reduce the extended configuration space to the one of (length) metric degrees of freedom. Part of these simplicity constraints are, however, second class \cite{Dittrich:2008ar,Perez:2012wv}. This prevents a sharp implementation of this part of the constraints, which is therefore imposed only weakly, but as strongly as allowed by the non-commutativity of the constraints. This non-commutativity is parametrized by the Barbero-Immirzi parameter \cite{BarberoG:1993xfb,Immirzi:1992ar}, which acts therefore as anomaly parameter, controlling fluctuations away from (length) metricity \cite{Dittrich:2012rj}.

As mentioned above, in effective spin foams the bulk intertwiner degrees of freedom are integrated out. This leaves us with the second class part of the constraints in terms of the area variables only.\footnote{That the constraints are second class follows also from the discreteness of the area spectra and the polynomial nature of the constraints. In terms of the area eigenvalues, the constraints appear as diophantine equations, which have too few solutions to allow for a semi-classical regime \cite{Asante:2020qpa}.} The constraints can be localized to pairs of neighbouring 4-simplices, and can thus be associated to the tetrahedra shared by these pairs. For each pair they force the area variables associated to its 16 triangles to arise from a consistent length assignment to its 14 edges. Weak implementation of these constraints leads to Gaussian factors $G_\tau$, whose variances are determined by the anomaly. 

These Gaussian factors make up one part of the effective spin foam amplitudes. The other part is given by the exponential\footnote{
Other choices, like the cosine of the Regge action, which implements a summation over orientations of the 4-simplices, are also possible \cite{Christodoulou:2012af, Engle:2011un}
}
of the Area Regge action $S_A$.
In summary, the effective spin foam amplitudes take the simple form 
\begin{equation}\label{esf}
{ Z} = \sum_{\{a\}} \mu(a) \,  e^{\imath S_A(a)} \,\prod_\tau G_\tau(a)  
\end{equation} 
where $\mu(a)$ is a measure factor to be fixed, for example, by demanding discretization invariance \cite{Dittrich:2011vz}.

Similar to the construction of previous spin foam models from gauge theoretic topological quantum field theories, effective spin foams  can be motivated from higher gauge theory, where the exponential of the Regge action appears directly (and not only as semi-classical limit) as an amplitude of a related topological quantum field theory \cite{Baratin:2014era,Mikovic:2011si,Asante:2019lki}.
Note that the EPRL/FK models \cite{Engle:2007wy,Freidel:2007py} can be cast into a form similar to \eqref{esf}, that is as state-sums of products of oscillating vertex amplitudes and products of terms enforcing gluing conditions between vertices \cite{Asante:2022lnp}.

The numerical accessibility of effective spin foams has been leveraged to elucidate the so-called flatness problem \cite{Bonzom:2009hw,Hellmann:2013gva}. Their transparent form of the dynamics has allowed a perturbative analysis of the refinement limit \cite{Dittrich:2022yoo}. In the following, we will shortly review these results.

\subsection{Discrete gravity dynamics from semi-classical limit}

The study of the semi-classical limit of spin foams \cite{Barrett:1998gs,Conrady:2008mk,Barrett:2009gg,Han:2011re}, i.e., of the limit $\hbar \rightarrow 0$\footnote{For spin foams, this is the limit of large areas associated to the triangles in the triangulation.}, has revealed that configurations with non-vanishing curvature are suppressed \cite{Hellmann:2013gva}.  This suggested that spin foams might not lead to general relativity, and became known as `flatness problem' \cite{Bonzom:2009hw,Hellmann:2013gva,Engle:2020ffj,Oliveira:2017osu,Dona:2020tvv,Gozzini:2021kbt}.

Although the flatness problem has been first identified for EPRL/FK models, it has been shown,  that it is a generic\footnote{A second more technical requirement is that the action is not stationary along the gradient of the constraints. Thus, examples where second class constraints are used for gauge fixing constant directions in the action avoid this problem.} feature for path integrals with weakly imposed constraints \cite{Asante:2020qpa,Asante:2020iwm}. As the need for weakly imposed constraints follows from the discreteness of the area spectra \cite{Asante:2020qpa}, one has to expect a version of the flatness problem for all models with discrete (and approximately equidistant) area spectra.

But with the insight that the flatness problem results from an anomalous constraint algebra, arises a resolution, namely that together with the $\hbar \rightarrow 0$ limit, one also has to take the anomaly parameter, thus the Barbero-Immirzi parameter, to be small \cite{Han:2013ina,Asante:2020iwm}.  Now, as the Barbero-Immirzi parameter does controls the spectral gap for spatial areas \cite{Rovelli:1994ge, Ashtekar:1996eg, Wieland:2017cmf}, one would like to know, how small the Barbero-Immirzi parameter has to be, in order to allow for a semi-classical regime with gravitational equations of motion.


The computational advantages of effective spin foams allowed the first explicit computation  \cite{Asante:2020iwm} of geometric expectation values for a triangulation, sufficiently large to test the equations of motion.
The results have shown that (effective) spin foams do implement a discrete version of the Einstein equations. The range of allowed parameters is larger than suggested by simple scaling arguments \cite{Han:2013ina,Asante:2020iwm}, and includes, even for large curvature of the classical solutions, Barbero-Immirzi parameters as large as $\sim 0.1-0.2$. This range includes values, which are suggested by black hole state counting \cite{Ashtekar:1997yu, Agullo:2008yv,Engle:2011vf,BarberoG:2015xcq}.

This has resolved the flatness problem for small triangulations. One might however ask, whether the flatness problem reappears for triangulations with many building blocks. Considering larger and larger triangulations, we have more and more degrees of freedom describing fluctuations away from length metric configurations. These might turn out to dominate in the refinement limit.

\subsection{The perturbative refinement limit of spin foams}


Ignoring the measure term, we can extract from the effective spin foam amplitude (\ref{esf}) an action
\ba\label{esfa}
S=S_A(a) -\imath \sum_\tau \ln G_\tau(a) \q .
\ea
The works \cite{Dittrich:2021kzs,Dittrich:2022yoo} assume that one can apply the background field method to effective spin foams, and therefore considers an expansion of the action (\ref{esfa}) around a flat solution. More precisely, one considers  triangulations with the symmetries of the hypercubical lattice, embedded into flat space. This embedding of a hypercubical lattice comes with two parameters, namely the lattice constant (defined by the background lengths of the edges of the hypercubes), and the number $N^4$ of hypercubes in the lattice. Choosing $N$ large, we can consider fluctuations with a large wavelength as compared to the lattice constant. Truncating the action to quadratic order in the fluctuations, we can apply a lattice Fourier transform, and fluctuations of different wavelengths will not interact with each other.

In the refinement (or continuum) limit we consider the dynamics at wavelengths much larger than the lattice constant. Here it is important to identify terms in the action, which have a homogeneous scaling in the lattice constant. Such terms can be identified after applying a variable transformation to a set of variables that describe an area metric. An area metric \cite{Schuller:2005yt} measures the areas of parallelograms in the tangent space. It has 20 components as compared to the 10 components of the length metric. One can however define a length metric from an area metric \cite{Borissova:2022clg}, and thus split the area metric into length metric and non-length-metric degrees of freedom. 

The action for the length metric degrees of freedom is of zeroth order in the lattice constant and provides the dominant part in the continuum limit. This action agrees with the linearized Einstein-Hilbert action. Integrating out the non-length-metric degrees of freedom, one finds a subleading correction (at quadratic order in the lattice constant) to the Einstein-Hilbert action \cite{Dittrich:2022yoo}. This correction is given by the square of the (linearized) Weyl curvature, and arises because of the extension of the configuration space from length to area variables.

This behaviour is explained by the fact that the length metric degrees of freedom are massless (the part of the action quadratic in these variables is also quadratic in derivatives), whereas the non-length-metric degrees of freedom come with a mass, determined by the lattice constant. Thus, these non-length-metric degrees of freedom essentially localize onto regions with size of a few lattice constants. Surprisingly, the dynamics of the Area Regge action itself provides such mass terms --- one actually does not need the constraint terms in (\ref{esfa}). Adding these constraint terms, one does affect the mass terms, but it does not change the structure of the results \cite{Dittrich:2021kzs}. 

As the various spin foam models differ in the details of their constraint implementation, these results hint at a universality mechanism that arises in the continuum limit. This is confirmed by an analysis that starts directly with the continuum field theory, underlying spin foam construction, namely the Plebanski action \cite{Plebanski:1977zz}.  The weak implementation of the constraints can be modelled by a modified Plebanski action \cite{Krasnov:2008fm,Krasnov:2009iy}, where the constraints are replaced by mass terms. Using this approach one can model the extension of the length metric configuration space to an area metric configuration space \cite{Borissova:2022clg}. Integrating out the non-length-metric degrees of freedom one finds again the Einstein-Hilbert action with a correction, given by the square of the Weyl curvature \cite{Borissova:2022clg}.

To summarize, effective spin foams have in particular allowed to probe the aspects of spin foam dynamics, which arise from the extension of the configuration space of length (metric) variables. Applying a `naive' semi-classical limit $\hbar \rightarrow 0$, this extension leads to the so-called flatness problem. This issue can be resolved by recognizing that the Barbero-Immirzi parameter is an anomaly parameter and has to be chosen to scale with $\hbar$ for a proper semi-classical limit. Explicit computations of expectation values in effective spin foams have identified semi-classical regimes with a gravitational dynamics, allowing for finite small values of $\hbar$ and the Barbero-Immirzi parameter \cite{Asante:2020iwm}. One can furthermore consider a perturbative continuum limit \cite{Dittrich:2021kzs,Dittrich:2022yoo}. This reveals that all the non-length metric degrees are massive, whereas the length metric degrees of freedom are massless. Thus, in the continuum limit, the non-length-metric degrees of freedom are suppressed. This provides a mechanism for resolving the flatness problem in the continuum limit, which works for arbitrary values of the Barbero-Immirzi parameter.

The lattice techniques developed in \cite{Dittrich:2021kzs,Dittrich:2022yoo} may also allow to extract a renormalization flow for the perturbative effective action, along the lines of \cite{Bahr:2010cq}.

\section{Concluding remarks}\label{Sec:Schluss}

Gravitational spin foams are defined on triangulations, which act as regulator in their path integral construction. This makes spin foams regulator (that is triangulation) dependent. It furthermore breaks diffeomorphism symmetry and introduces many unwanted ambiguities. We have argued that all these issues can be resolved, if one constructs improved and perfect actions. Such actions lead however to non-local couplings, which are difficult to handle in particular for the spin foam framework. Non-local couplings are avoided in the consistent boundary formalism. It provides a renormalization framework that serves to construct consistent (that is regulator independent) amplitudes in background independent theories such as spin foams, for spacetime regions with more and more complex boundary data. To implement this in practice one has to choose truncations. Importantly the consistent boundary framework allows to identify truncations that are adapted to the dynamics such that they minimize the induced error for the evaluation of the partition function.  Tensor network coarse graining provides concrete algorithms that can be used to implement the consistent boundary framework. Originally constructed for 2D systems, they needed to be improved and modified to allow applications to higher dimensional systems with gauge symmetries. In particular, decorated tensor networks have been successfully applied to 3D non-Abelian lattice gauge systems, emulating spin foams. 

Applying such coarse graining algorithms to the 4D gravitational spin foam models remains an enormous computational challenge. This is in no small part due to the complexity of the spin foam amplitudes. We presented two approaches that lead to a simplification of the amplitudes (which can also be combined). Restricted spin foams combine a symmetry reduction of degrees of freedom with semi-classical approximations. This has allowed to derive a renormalization flow, which illustrated the restoration of a partial set of diffeomorphism symmetries at the fixed point. Effective spin foams are much more accessible to numerical computations than traditional models and offer a transparent encoding of the dynamics. These features can be leveraged to show that the flatness problem, which seemed to endanger a suitable semi-classical limit for spin foams, can be resolved both on the discrete level and in the refinement limit.  

In particular, effective spin foams allowed for a perturbative refinement limit. This is described by the Einstein-Hilbert action and a Weyl squared curvature term. The latter arises from a key feature of spin foams, namely the extension of the underlying configuration space from length to area metrics. Although this extension is forced from the discrete area spectra of loop quantum gravity, we should note that the found action does not capture directly the effects resulting from these discrete spectra. 

In closing, we remark that many questions about the non-perturbative refinement limit remain open. Spin foams feature many degenerate configurations, e.g. so-called vector geometries \cite{Conrady:2008mk, Barrett:2009gg, Dona:2017dvf, Dona:2019dkf}, that might turn out to dominate in the limit. This holds even more so for Lorentzian signature models \cite{Kaminski:2017eew, Liu:2018gfc, Simao:2021qno}. In addition, Lorentzian models can also feature topology change, e.g. generation of baby universes \cite{Asante:2021zzh, Asante:2021phx}.  Luckily, effective spin foams allow for an easier control of such configurations. 

Many non-perturbative (Euclidean) lattice approaches suffer from the conformal factor problem \cite{Loll:1998aj}. This leads, in e.g. Regge calculus, to an abundance of spike configurations \cite{Ambjorn:1997ub}, that is vertices, where all adjacent edges have (arbitrarily) large lengths. A related problem for spin foams are bubble divergences \cite{Perini:2008pd, Riello:2013bzw, Bonzom:2013ofa,Chen:2016aag,Dona:2018pxq}. To achieve a suitable non-perturbative refinement limit, it will be essential to control such divergences. A first step would be to show that spike configurations can be dealt with in Lorentzian quantum Regge calculus. 

\begin{acknowledgement}
BD and SSt appreciate deeply many discussions and collaborations with Benjamin Bahr on the topics discussed here.
Research at Perimeter Institute is supported in part by the Government of Canada through the Department of Innovation, Science and Economic Development Canada and by the Province of Ontario through the Ministry of Colleges and Universities.
SKA is supported by the Alexander von Humboldt foundation.
SSt gratefully acknowledges support by the Deutsche Forschungsgemeinschaft (DFG, German Research Foundation) - Projektnummer/project-number 422809950. 
\end{acknowledgement}

\bibliographystyle{utphys}
\bibliography{bibliography}

\end{document}